\documentclass[a4paper,10pt]{article}
\usepackage{graphicx}
\renewcommand{\theequation}{\arabic{section}.\arabic{equation}}
\newcommand{\beq}{\begin{equation}}
\newcommand{\enq}{\end{equation}}

\newcommand{\mapright}[1]{%
\smash{\mathop{%
\hbox to 1.0cm{\rightarrowfill}}\limits^{#1}}}
\newcommand{\mapleft}[1]{%
\smash{\mathop{%
\hbox to 1.3cm{\leftarrowfill}}\limits^{#1}}}

\newcommand{\no}{\nonumber}
\newtheorem{lem}{Lemma}
\newtheorem{thm}{Theorem}
\newtheorem{conj}{Conjecture}

\newtheorem{defi}{Definition}

\newcommand{\beqy}{\begin{eqnarray}}
\newcommand{\enqy}{\end{eqnarray}}

\begin{document}

\begin{titlepage}
\vglue 3cm

\begin{center}
\vglue 0.5cm
{\Large\bf $O(-2)$ Blow-up Formula via Instanton Calculus
on $\widehat{{\bf C}^2/{\bf Z}_2}$ and Weil Conjecture}
\vglue 1cm
{\large Toru Sasaki}
\vglue 0.5cm
{\it Graduate School of Mathematics,
Nagoya University, Nagoya, 464-8602, Japan}\\
{x03001e@math.nagoya-u.ac.jp}

\baselineskip=12pt

\vglue 1cm
\begin{abstract}
We calculate Betti numbers of the framed moduli space of instantons
on $\widehat{{\bf C}^2/{\bf Z}_2}$,
under the assumption that the corresponding torsion free sheaves
$E$ have vanishing properties($Hom(E,E(-l_\infty))=Ext^2(E,E(-l_\infty))=0$).
Moreover we derive the generating function of Betti numbers
and obtain closed formulas.
On the other hand, we derive a universal relation between
the generating function of Betti numbers of 
the moduli spaces of stable sheaves on $X$
with an $A_1$-singularity and that on $\hat{X}$ blow-uped at the singularity,
by using Weil conjecture.
We call this the $O(-2)$ blow-up formula.
Applying this to $X={\bf C}^2/{\bf Z}_2$ case, we reproduce the formula
given by instanton calculus.
\end{abstract}
\end{center}
\end{titlepage}

\section{Introduction}
\setcounter{equation}{0}
In the recent years, instanton calculus has been
used to determine the non-perturbative effects of supersymmetric theories
exactly, which are interpreted as prepotentials or the partition functions
\cite{Nek,N-O,Fucito,po}.
Key point of these successes is owing to localization theorem.
We apply this instanton calculus
to determine the twisted ${\cal N}=4~SU(2)$ partition function
on $\widehat{{\bf C}^2/{\bf Z}_2}$
and the generating function of Poincar\'e polynomials
of instanton moduli spaces for rank two.
These partition function and generating function 
of Poincar\'e polynomials
are written by  beautiful closed formulas as follows.
For the framed moduli spaces of instantons
with rank 2 and Chern classes $c_1=0,c_2 = n$ on $\widehat{{\bf C}^2/{\bf Z}_2}$,
the generating function of Poincar\'e polynomials of them is given by
\begin{thm}
\beqy
&&
\sum_{n\in\frac{\bf Z}{2}}^\infty P_t(\hat{M}(2,0,n))q^n
\no
\\
&=&
\prod_{d=1}^\infty\frac{1}{(1-t^{4d}q^d)(1-t^{4d-2}q^d)^2(1-t^{4d-4}q^d)}
\prod_{d=1}^\infty\frac{1-(-t^2q^{\frac{1}{2}})^dt^{-2}}{1+(-t^2q^{\frac{1}{2}})^d}.
\enqy
\end{thm}
On the other hand, we derive the above formula in the different way.
Let $X$ be a complex surface with an $A_1$-singularity,
and $\hat{X}$ be the surface blow-uped at the singularity.
By using Weil conjecture and elementary transformations,
we derive a universal relation between
the generating function of Betti numbers of the moduli spaces
of stable sheaves on $X$ and that on $\hat{X}$, 
called as the ${\cal O}(-2)$ blow-up formula.
After deriving the generating function of Betti numbers of the moduli spaces
of stable sheaves on ${\bf C}^2/{\bf Z}_2$,
we reproduce (1.1) by using the $O(-2)$ blow-up formula.

In \cite{jin,jin2}, we used the $O(-2)$ blow-up formula
in order to verify the equivalence
between the twisted ${\cal N}= 4$ partition function on orbifold $T^4/{\bf Z}_2$
and that on $K3$.
It is well-known that $K3$ surface
is constructed by orbifold $T^4/{\bf Z}_2$ \cite{Fukaya}.
First one divides $T^4$ surface by ${\bf Z}_2$.
Then 16 singularities appear.
So one makes a smooth surface from this singular surface
by blowing up these 16 singularities.
The resulting smooth surface is Kummer surface
(a kind of $K3$ surface).
We tried to reconstruct the geometrical processes
in the partition function level \cite{jin, jin2}.
In the above processes,
the singular surface $T^4/{\bf Z}_2$ (the simply divided surface by ${\bf Z}_2$)is denoted by $S_0$ and called the contribution from $S_0$ as that from the untwisted sector.
Blowing up a singularity,
we obtain the contribution from the blow-up process,
and call them a blow-up formula.
When we blow up 16 singularities in $S_0$,
we obtain 16th power of the blow-up formula
and call them as the contribution from the twisted sector.
We combine the untwisted sector with the twisted sector,
so that the total partition function reproduces the modular property of ${\cal N}=4$
on $K3$ \cite{M-O}.
Finally we obtain the same partition function as that on $K3$
given by Vafa and Witten \cite{vafa-witten}.
However the blow-up formula used above is not rigorous mathematically.
There are two evidences of the justification.
First, the ${\cal O}(-1)$ blow-up formula,
which was introduced by Yoshioka et al.\cite{yoshioka,Li-Qin,Li-Qin2}
has the form $\theta^{(1)}_{A_{n-1}}(\tau)/{\eta(\tau)^n}$ for $SU(n)$.
Here $\theta^{(1)}_{A_{n-1}}(\tau)$ and ${\eta(\tau)}$
are a level $1~A_{n-1}$ theta function and Dedekind's eta function respectively
(see Appendix A).
In \cite{yoshioka}, Yoshioka calculated the effect of the blowing up process:
$\hat{{\bf P}^2} \to {\bf P}^2$ to the partition function.
Here $\hat{{\bf P}^2}$ stands for one point blowing up at the origin.
The ${\cal O}(-1)$ blow-up formula is given by the ratio of the partition function
on $\hat{{\bf P}^2}$ and that on ${\bf P}^2$.
For the ${\cal O}(-2)$ blow-up formula,
we replace $\theta^{(1)}_{A_{n-1}}(\tau)$
by $\theta^{(2)}_{A_{n-1}}(\tau)$,
since we use an $O(-2)$ curve in blowing up a singularity in $S_0$.
Secondary, we have another evidence from a stringy picture.
The ${\cal N}=4~SU(n)$ super Yang-Mills theory on $K3$
can be derived from a IIA string compactified
on $T^2\times K3\times ALE_{A_{n-1}}$\cite{jin3}.
Here $ALE_{A_{n-1}}=\widehat{{\bf C}^2/{\bf Z}_n}$.
On the other hand,
the ${\cal N}=4~U(1)^{24}$ super Yang-Mills theory on $ALE_{A_{n-1}}$
can be derived from a heterotic string compactified
on $T^2\times T^4\times ALE_{A_{n-1}}$.
These two ${\cal N}=4$ theories are connected by hetero-IIA string duality\cite{hj}.
The ${\cal N}=4~U(1)$ partition function on $ALE_{A_{n-1}}$
was already determined by Nakajima\cite{naka}.
It has the same form as a blow-up formula as expected.
The twisted ${\cal N}=4$ partition function is
given by the generating function of Euler numbers of instanton moduli spaces
\cite{vafa-witten,laba},
which can be considered as a special case of the generating function of
Poincar\'e polynomials with parameter $t=-1$.
Thus, if we can calculate the generating function of Poincar\'e polynomials
of instanton moduli spaces, we easily obtain
the twisted ${\cal N}=4$ partition function.
So, we concentrate on deriving the generating function for 
the ${\cal O}(-2)$ blow-up formula.

Here we briefly explain the blow-up formula
for the $S$-duality conjecture of Vafa-Witten\cite{vafa-witten}.
We do not consider the conventional blow-up formula
given by Fintushel and Sterm\cite{F-S} in this article.
Roughly speaking, the blow-up formula
is the universal relation between
the Euler numbers( or the generating function of the Betti numbers) 
of instanton moduli spaces on a smooth four manifold
and those on the blow-up of the smooth four manifold.
The universal relation is independent of the four manifold.
Yoshioka derived this formula
by using Weil conjecture and elementary transformations\cite{yoshioka}.
Li and Qin also derived this formula
by using virtual Hodge polynomials\cite{D-K}
in the same way\cite{Li-Qin,Li-Qin2}.
Kapranov derived the similar formula
corresponding to an ${\cal O}(-d)$ blow-up formula
in the connection with Kac-Moody algebra\cite{kap}. 

Nakajima and Yoshioka studied the framed moduli space
of instantons on $\hat{{\bf C}^2}$ 
by using the technique of instanton calculus
\cite{naka,n-y1,n-y2}.
In \cite{naka}, they reproduce the ${\cal O}(-1)$
blow-up formula derived in the different way\cite{yoshioka,Li-Qin,Li-Qin2}.
Let $F_2$ be a Hirebruch surface which has an ${\cal O}(-2)$ curve\cite{ba}.
Considering a framed moduli space of torsion free sheaves $E$ on $F_2$,
we derive Betti numbers of the framed moduli spaces 
of instantons on $\widehat{{\bf C}^2/{\bf Z}_2}$.
Moreover we obtain the generation function of them
as a beautiful closed formula (1.1).
This formula coincides with the result given by T.Hausel\cite{hausel}.
In the calculation, we assume vanishing properties of the torsion free sheaves
($Hom(E,E(-l_\infty))=Ext^2(E,E(-l_\infty))=0$).
We introduce fractional line bundles
which are not defined rigorously,
since we want to see the relation between level $2$ theta functions
and the ${\cal O}(-2)$ curve blow-up formula(See also Sec.A.2).
To derive the ${\cal O}(-2)$ blow-up formula,
we have to determine Betti numbers of the moduli spaces
of instantons on ${\bf C}^2/{\bf Z}_2$.
However we cannot determine this by using the technique of instanton calculus.
Thus we do in the different way.

Yoshioka derived the ${\cal O}(-1)$ blow-up formula
by using Weil conjecture and elementary transformations\cite{yoshioka}.
We want to generalize his methods to the ${\cal O}(-2)$ case.
The most difficult point is the precise definition of
the stable vector bundle on a singular surface.
However we avoid this difficult point 
and consider the formal generalization 
of the ${\cal O}(-1)$ case to the ${\cal O}(-2)$ case.
Here we think that the Weil conjecture is valid on a singular surface
with an $A_1$-singularity and use this conjecture on the surface.
Under the assumption of the existence of the stable vector bundles,
we obtain a universal relation between
the generating function of Betti numbers of
the moduli spaces of stable sheaves on $X$
with an $A_1$-singularity and that on $\hat{X}$ blow-uped at the singularity.
We check this ${\cal O}(-2)$ blow-up formula
for $X={\bf C}^2/{\bf Z}_2$ case,
and obtain the same result as one given by instanton calculus.
The generating function of Betti numbers of
the moduli spaces of stable sheaves on ${\bf C}^2/{\bf Z}_2$ 
is obtained as follows.
We separate the contribution on $\widehat{{\bf C}^2/{\bf Z}_2}$
into that from the untwisted sector and that from the twisted sector.
Then, the contribution from the untwisted sector
can be obtained by taking ${\bf Z}_2$-invariant part of
the contribution on ${\bf C}^2$.

The organization of the article is as follows.
In Sec.2, we introduce the blow-up and Weil conjecture used below.
In Sec.3.1, we introduce the framed moduli space of torsion free sheaves
on ${\bf P}^2(F_2)$
and that of instanton on $S^4(\widehat{{\bf C}^2/{\bf Z}_2 })$.
In Sec.3.2, we study torus actions and their fixed point set.
In Sec.3.3, we calculate Betti numbers of the framed moduli space of 
torsion free sheaves on ${\bf P}^2(F_2)$.
In Sec.4.1, we calculate Betti numbers of the moduli space of
stable vector bundles on ${\hat X}(X)$.
In Sec.4.2, we provide the useful formulas
to count Betti numbers of the moduli spaces of
stable sheaves of non-vector bundle on ${\hat X}(X)$.
Then, we derive the ${\cal O}(-2)$ blow-up formula.
In Sec.4.3, we derive Betti numbers of the moduli spaces of
stable sheaves on ${\bf C}^2/{\bf Z}_2$.
Then, we reproduce the formula given in Sec.3.3
by using the ${\cal O}(-2)$ blow-up formula.
In Sec.5, we summarize our results
and make some comments on the other related works.
In Sec.A.1, we introduce level $l$ theta functions used in this article.
In Sec.A.2, we show that level $2$ theta functions appear naturally in instanton calculus on ${\widehat{{\bf C}^2/{\bf Z}_2}}$.
In Sec.B, we verify an identity appeared in derivation of (1.1). 
In Sec.C, we calculate Betti numbers of the framed moduli spaces
for odd $c_1$ under some assumptions.

\section{Preliminary}
\setcounter{equation}{0}
\subsection{Blow-up}
Following \cite{Fukaya}, we introduce the blow-up:
\begin{defi}
Under the immersion
\beq
I:{\bf C}^2 -\{ \vec{0} \} \to {\bf C}^2\times {\bf P}^1,
\enq
the closure of the image is denoted by ${\hat {{\bf C}^2}}$.
We call  ${\hat {{\bf C}^2}}$ as the blow-up of ${\bf C}^2$ at $\{\vec{0}\}$.
\end{defi}

We introduce an open subset of ${\bf C}^2$ as
\beq
D_C^2=\{ (z_1,z_2)\in {\bf C}^2| |z_1|^2+|z_2|^2<1 \}.
\enq
The whole complex linear maps
preserving inner product are denoted by $U(2)$
and the finite subgroup is denoted by $\Gamma$.
We consider the immersion
\beq
I:D_C^2 -\{ \vec{0} \} \to D_C^2\times {\bf P}^1.
\enq
Then $\Gamma$-action is preserved. So, we can also define
the following blow-up:
\begin{defi}
\beq
{\bar I}:(D_C^2 -\{ \vec{0} \} )/\Gamma\to D_C^2\times {\bf P}^1/\Gamma,
\enq
the closure of the image is denoted by ${\widehat {D_C^2/\Gamma}}$.
We call  ${\widehat {D_C^2/\Gamma}}$
as the blow-up of ${D_C^2/\Gamma}$ at $\{\vec{0}\}$.
\end{defi}
In $\Gamma={\bf Z}_r$ case, we consider that ${D_C^2/{\bf Z}_{r+1}}$ has 
an $A_r$-singularity.
\subsection{Weil Conjectures}
Let ${\bf F}_q$ be a finite field with $q$ elements,
$\bar{\bf F}_q$ the algebraic closure of ${\bf F}_q$.
For a scheme $X$ over ${\bf F}_q$,
$\bar{X}$ denotes $X\otimes_{{\bf F}_q}\bar{\bf F}_q$
and $X({\bf F}_q)$ the set of ${\bf F}_q$-rational points of $X$.
For a smooth projective variety $X$ over ${\bf C}$,
$b_i(X)$ is the $i$-th Betti number of $X$,
$\chi(X):=\sum_i(-1)^ib_i(X)$ the Euler number of $X$, and
\beq
P_t(X):=\sum_ib_i(X)t^i
\enq
the Poincar\'e polynomial of $X$.
The zeta function of $X$ over ${\bf F}_q$ is defined as
\beq
Z_q(X,t):=\exp(\sum_{r>0}(\# X({\bf F}_{q^r})\frac{t^r}{r})).
\enq
\begin{thm}[Deline]\cite{Deligne}\\
Let $X$ be a smooth projective variety of dimension $n$ over ${\bf F}_q$.
\begin{itemize}
\item[(1)]$Z_q(X,t)$ is a rational function on $t$.
\item[(2)]
\beq
Z_q(X,t)=\prod_{i=0}^{2n}P_i(X,t)^{i+1},
\enq
where $P_i(X,t)=\prod_{j=1}^{b_i}(1-\alpha_{i,j}t), |\alpha_{i,j}|=q^{i/2}.$
\item[(3)]
We have \beq
Z_q(X,1/q^nt)=\pm q^{e(\bar{X})/2}t^{e(\bar{X})}Z_q(X,t),
\enq
where $e(\bar{X})=\sum_i(-1)^idegP_i$.
\item[(4)]
If $X$ is a good reduction of a smooth projective variety $Y$ over ${\bf C}$,
then $b_i=b_i(Y)$.
\end{itemize}
\end{thm}

From this theorem, replacing $\alpha_{i,j}$ by $(-z)^i$,
we obtain the Poincar\'e polynomial of $Y$:
\beq
\# X({\bf F}_q)=\sum_{i,j}(-1)^i\alpha_{i,j}\to P_z(Y)=\sum_ib_i(Y)z^i.
\enq
We consider the case of $X={\bf P}^n$ for an example.
By using the Frobenius morphism $F:X_{\bar{\bf F}_q}\to X_{\bar{\bf F}_q}$ sending
$(z_0:z_1:\cdots :z_n)$ to $(z_0^q:z_1^q:\cdots :z_n^q)$,
we obtain the fixed point set
$\{ x\in X_{\bar {\bf F}_q}| F(x)=x \} = X({\bf F}_q)$.
By Lefshetz fixed point theorem,
$\# X({\bf F}_q)=\sum_i(-1)^i tr(F_i^*)=\sum_i^nq^i,$
where $F_i^*:H^i(X)\to H^i(X)$ is the endomorphism induced by $F$.
Thus we obtain
\beq
\# {\bf P}^n({\bf F}_q)=\sum_{i=0}^nq^i{\sim\atop\to} P_z({\bf P}^n)=\sum_{i=0}^nz^{2i},
\enq
where we set $q=z^2$.

Although the Weil conjecture is true on a smooth projective variety,
we apply this to the case of $X$= a singular surface
with an $A_r$-singularity in this article.

\section{Instanton Calculus on $\widehat{{\bf C}^2/{\bf Z}_2}$}
\setcounter{equation}{0}

\subsection{Framed Moduli Space}
The framed moduli space is mainly defined  in the two ways
as that of instanton and that of torsion free sheaf.
At first, we think of the framed moduli space
of instantons on $S^4$ and that of torsion free sheaves on ${\bf P}^2$
\cite{n-y}.

We introduce the hyper-K\"ahler geometry used below\cite{nak}.
\begin{defi}
Let $X$ be a $4n$-dimensinal manifold.
A hyper-K\"ahler structure of $X$
consists of a Riemanian metric $g$
and a triple of almost complex structures ${\cal I,J,K}$
which satisfy the following conditions:
\begin{itemize}
\item[(1)]
\beq
g({\cal I}v,{\cal I}w)=g({\cal J}v,{\cal J}w)=g({\cal K}v,{\cal K}w)=g(v,w) \mbox{ for } v,w\in TX.
\enq
\item[(2)]
$({\cal I,J,K}) $ satisfies a relation
\beq
{\cal I}^2={\cal J}^2={\cal K}={\cal IJK}=-1.
\enq
\item[(3)]
${\cal I,J}$ and ${\cal K}$ are parallel with respect to the Levi-Civita connection of $g$,
\beq
\nabla {\cal I}=\nabla {\cal J}=\nabla {\cal K}=0.
\enq
\item[(4)]
For ${\cal I,J}$ and ${\cal K}$, 2-forms $\omega_1,\omega_2$ and $\omega_3$ are defined
by
\beq
\omega_1(v,w):=g({\cal I}v,w),\omega_2(v,w):=g({\cal J}v,w),\omega_3(v,w):=g({\cal K}v,w) \mbox{ for } v,w\in TX,
\enq
which satisfy $d\omega_1=d\omega_2=d\omega_3=0$.
(These $\omega_1,\omega_2$ and $\omega_3$ are called the K\"ahler forms
associated with $(g,{\cal I}),(g,{\cal J})$ and $(g,{\cal K})$, respectively.)
\end{itemize}
\end{defi}
Suppose that a compact Lie group $G$  acts on $X$
preserving $g,{\cal I,J,K}$.
The Lie algebra of $G$ is denoted by ${\cal G}$,
and its dual is denoted by ${\cal G}^*$.
\begin{defi}
A map
\beq
\mu=(\mu_1,\mu_2,\mu_3):X\to {\bf R}^3\otimes {\cal G}^*
\enq
is said to be a hyper-K\"ahler momentum map
if it satisfies the following conditions:
\begin{itemize}
\item[(1)]
$\mu$ is $G$-equivalent,i.e. $\mu(g\cdot x)=Ad_{g^{-1}}^*\mu(x).$
\item[(2)]
$<d\mu_i(v),\xi>=\omega_i(\xi^*,v)$
for any $v\in TX$, any $A\in {\cal G}$ and $i=1,2,3,$
where $\xi^*$ is a vector field generated by $\xi$.
\end{itemize}
\end{defi}

Let $V,W$ be hermitian vector spaces
whose dimensions are $n,r$,respectively.
For these spaces, we define a complex vector space ${\bf M}(r,n)$
as
\beq
{\bf M}(r,n):=\{(B_1,B_2,I,J)| B_1,B_2\in Hom(V,V),I\in Hom(W,V),J\in Hom(V,W) \}.
\enq
We consider an action of $g\in U(V)$ on ${\bf M}(r,n)$  given by  
\beq
{\bf M}(r,n)\in (B_1,B_2,I,J) \to (g^{-1}B_1g,g^{-1}B_2g,g^{-1}I,Jg).
\enq
We define a momentum map $\mu_{1}: {\bf M}(r,n)\to U(V)$,
\beq
\mu_{1}(B_1,B_2,I,J):=\frac{i}{2}([B_1,B_1^\dagger]+[B_2,B_2^\dagger]+II^\dagger-J^\dagger J).
\enq
We also define a momentum map $\mu_{\bf C}: {\bf M}(r,n)\to End(V)$,
\beq
\mu_{\bf C}(B_1,B_2,I,J):=[B_1,B_2]+IJ.
\enq
Note that $\mu_{\bf C}=\mu_2+i\mu_3$ in this case.

\begin{defi}
The framed moduli space of instantons on $S^4$
with rank $r$ and second Chern class $n$
is defined by
\beq
M(r,n):=\mu_{1}^{-1}(\frac{i}{2}\zeta id)\cap\mu_{\bf C}^{-1}(0)/U(V),
\enq
where $\zeta$ is a fixed positive real number.
\end{defi}
This space is known to be non-singular of dimension $2nr$.
$\mu_{1}^{-1}(\cdots)\cap\mu_{\bf C}^{-1}(\cdots)/U(V)$ stands for
a hyper-K\"ahler quotient.
We also give another type of description
of the framed moduli space
of instanton on $S^4$:
\begin{lem}\cite{nak}
The framed moduli space of instantons on $S^4$
with rank $r$ and second Chern class $n$
is given by
\beq
M(r,n)=\{(B_1,B_2,I,J)\in \mu_{\bf C}^{-1}(0)|
 (B_1,B_2,I,J)\mbox{ is stable }\}/U(V),
\enq
where the stable $(B_1,B_2,I,J)$ has no subspace $S\subset V$
which satisfies
\beq
B_\alpha(S) \subset S \mbox{ and } Im (I) \subset S.
\enq
\end{lem}
We define the framed moduli space of ideal instantons
on $S^4$
with rank $r$ and second Chern class $n$:
\begin{defi}
The framed moduli space of ideal instanton on $S^4$
with rank $r$ and second Chern class $n$
is defined by
\beq
M_0(r,n):=\mu_{1}^{-1}(0)\cap\mu_{\bf C}^{-1}(0)/U(V).
\enq
\end{defi}
We also give the framed moduli space
of ideal instantons on $S^4$
differently:
\begin{lem}\cite{nak}
The framed moduli space of ideal instantons on $S^4$
with rank $r$ and second Chern class $n$
is given by
\beqy
M_0(r,n)&=&
\mu_{\bf C}^{-1}(0)//GL(V)
\no
\\
&=&
\mbox{the set of closed }GL(V)\mbox{-orbits in }\mu_{\bf C}^{-1}(0),
\enqy
where $//$ means the affine algebro-geometric quotient.
\end{lem}
Since this space has singularities, we take the non-singular locus 
defined by
\begin{defi}
\beq
M_0^{reg}(r,n):=\{
[(B_1,B_2,I,J)]\in M_0(r,n)|
\mbox{the stabilizer in }U(V)\mbox{ of } (B_1,B_2,I,J) \mbox{ is trivial}
\}.
\enq
\end{defi}
This space is identified with the moduli space of genuine instantons
on $S^4$.
$M_0^{reg}(r,n)$ and $M_0(r,n)$ is related by
\beq
M_0(r,n)=\coprod_{k=0}^{n}M_0^{reg}(r,n-k)\times S^k {\bf C}^2.
\enq
This means that $M_0(r,n)$ is an Uhlenbeck compactification of $M_0^{reg}$.
Here $S^n X$ stands for $n$-th symmetric product of $X$.

Here we mention the difference between the framed moduli space
of instantons and the moduli space of instantons 
by using analytic terms\cite{nak}.
First we show that
the moduli space of instantons on a 4-dimensional hyper-K\"ahler manifold $X$
 is  a hyper-K\"ahler quotient.
For a smooth vector bundle $E$ over $X$ with a Hermitian metric
and the space of metric connections ${\cal A}$ on $E$,
the tangent space $T_A{\cal A}$ at $A\in {\cal A}$ 
is identified with $T_A{\cal A}\cong \Omega^1({\cal U}(E))$.
For the tangent space $T_A{\cal A}$,
we have a natural $L^2$-metric
and almost complex structures on $T_A{\cal A}$
induced from those on $X$.
We also have the group of gauge transformations ${\cal G}$ acting on ${\cal A}$.Then, the hyper-K\"ahler momentum map of the action of ${\cal G}$ on ${\cal A}$
\beq
\mu=(\mu_1,\mu_2,\mu_3):{\cal A}\to {\bf R}^3\otimes Lie{\cal G}^*
\cong  {\bf R}^3\otimes \Omega^4({\cal U}(E))
\enq
is given by
\beq
\mu_i({\cal A})=F_A^+\wedge\omega_i\in \Omega^4({\cal U}(E))~~(i=1,2,3).
\enq
Here $F_A^+$ is a self-dual part of the curvature 2-form of $A$,
and $\omega_i$ is the K\"ahler form associated with the complex structures on $X$.We define the moduli space of instantons on $X$ by $\mu^{-1}(0)/{\cal G}$.
This construction works even in the case $X={\bf C}^2$.
The framed moduli space of instantons is considered as
a quotient by a group of gauge transformations 
converging to the identity at the end of $X$.
Thus in the case $X={\bf C}^2$
we consider one point compactification $S^4={\bf C^2}\cup\{ \infty \}$.
The difference between the framed moduli space
of instantons and the moduli space of instantons
is whether the corresponding vector bundles have 
a converging property for a group of gauge transformations or not.

To mention the relation between $M(r,n)$ and $M_0(r,n)$,
we define the framed moduli space of torsion free sheaves on ${\bf P}^2$,
which is an alternative definition of the framed moduli space.

Let $M(r,n)$ be the framed moduli space of torsion free sheaves on ${\bf P}^2$
with rank $r$ and $c_2=n$,
which parametrizes isomorphism classes of $(E,\Phi)$ such that
\begin{itemize}
\item[(1)]
$E$ is a torsion free sheaf of rank$(E)=r,<c_2(E),[{\bf P}^2]>=n$
which is locally free in a neighborhood of $l_\infty$,
\item[(2)]
$\Phi:E|_{l_\infty}\to {\cal O}_{l_\infty}^{\oplus r}$
is an isomorphism called `framing at infinity'.
\end{itemize}
Here $l_\infty =\{[0:z_1:z_2]\in {\bf P}^2 \} \subset {\bf P}^2$
is the line at infinity.
Note that the existence of a framing $\Phi$ implies $c_1(E)=0$.
The equivalence between the framed moduli spaces
of instanton on $S^4$ and that of torsion free sheaves on ${\bf P}^2$
is explained in detail\cite{nak}.
Using this definition, we denote the relation between $M(r,n)$ and $M_0(r,n)$
\begin{thm}\cite{n-y}
There is a projective morphism
\beq
\pi: M(r,n)\to M_0(r,n)
\enq
defined by
\beq
(E,\Phi)\to (E^{\vee\vee},\Phi),Supp(E^{\vee\vee}/E))
\in M_0^{reg}(r,n^\prime)\times S^{n-n^\prime}{\bf C}^2,
\enq
where $E^{\vee\vee}$ is the double dual of $E$ and Supp$(E^{\vee\vee}/E)$
is the support of $(E^{\vee\vee}/E)$ counted with multiplicities.
\end{thm}
This morphism $\pi$ is the Hilbert-Chow morphism.
As an example of the Hilbert-Chow morphism, we write a theorem:
\begin{thm}\cite{nak}
\beq
M(1,n)=({\bf C}^2)^{[n]},M_0(1,n)=S^n({\bf C}^2),
\enq
where $(X)^{[n]}$ stands for the Hilbert scheme of $n$ points
on surface $X$.
\end{thm}

Now we move to the framed moduli space of instanton
on $\widehat{{\bf C}^2/{\bf Z}_2}$.
The similar treatment is done in \cite{n-y}.
To describe the corresponding torsion free sheaves,
we introduce the following manifold $F_2$:
\begin{equation}
F_2=\{([z_0:z_1:z_2],[z:w])\in {\bf P}^2\times {\bf P}^1;z_1w^2=z_2z^2\}
\end{equation}
as a Hirzebruch surface\cite{ba}.
Let $p:F_2\to {\bf P}^2$ denote the projection to the first factor.
We denote the inverse image of $\{z_0=0\}\subset {\bf P}^2$
under $p:F_2\to {\bf P}^2$
by $l_\infty$
and denote the exceptional set $\{z_1=z_2=0\}$ by $C$.
Note that $F_2\setminus l_\infty=\widehat{{\bf C}^2/{\bf Z}_2}$
and the self-intersection $[C]^2=-2$.
We interpret that $p:F_2\to {\bf P}^2$ is a kind of blow-up,
but is not different from that in Sec.2.
The blow-up in Sec.2 is treated in the next section.

In the remaining part, ${\cal O}$ denotes
the structure sheaf of $F_2$,
${\cal O}(C)$ the line bundle associated with the divisor $C$,
${\cal O}(mC)$ its $m$th tensor product.

Let ${\hat M}(r,k,n)$ be the framed moduli space of torsion free sheaves
$(E,\Phi)$
on $F_2$ with rank 2,
$ <c_1(E),[ C ]>=-k$ and
$<c_2(E)-\frac{r-1}{2r}c_1(E)^2,[F_2] >= n $.
This space is non-singular of dimension $2nr$.

In the same way as the framed moduli space of torsion free sheaves on $\hat{ {\bf P}^2}$\cite{n-y},
we assume
\begin{conj}
There is a projective morphism
\beq
\hat{\pi}: \hat{M}(r,k,n)\to M_0(r,n-\frac{1}{4r}k(r-k))(0\le k <2r)
\enq
defined by
\beq
(E,\Phi)\to ((p_*E)^{\vee\vee},\Phi),Supp(p_*E^{\vee\vee}/p_*E)+Supp(R^1p_*E)).
\enq
\end{conj}
We will give an evidence of Conjecture 1 in Sec.3.3.
\subsection{Torus Actions and their Fixed Point Set}
Let us define an action of the $(r+2)$-dimensional torus ${\tilde T}={\bf C}^*\times {\bf C}^*\times T^r$
on ${M}(r,n)$\cite{n-y,n-y1}. 
For $(t_1,t_2)\in {\bf C}^*\times {\bf C}^*$,
let $F_{t_1,t_2}$ be an automorphism of ${\bf P}^2$
defined by
\begin{equation}
F_{t_1,t_2}([z_0:z_1:z_2])=([z_0:t_1z_1:t_2z_2]).
\end{equation}
For diag$(e_1,..,e_r)\in T^r$,
let $G_{e_1,..,e_r}$ denotes the isomorphism of ${\cal O}_{l_\infty}^{\oplus r}$ given by
\beq
{\cal O}_{l_\infty}^{\oplus r}\ni (s_1,..,s_r)\mapsto (e_1s_1,..,e_rs_r).
\enq
Then for $(E,\Phi)\in M(r,n)$, we define
\beq
(t_1,t_2,e_1,..,e_r)\cdot (E,\Phi)=((F_{t_1,t_2}^{-1})^*E,\Phi^\prime),
\enq
where $\Phi^\prime$ is the composite of homomorphisms
\beq
(F_{t_1,t_2}^{-1})^*E|_{l_\infty}
{(F_{t_1,t_2}^{-1})^*\Phi \atop \longrightarrow }
(F_{t_1,t_2}^{-1})^*{\cal O}_{l_\infty}^{\oplus r}
\to {\cal O}_{l_\infty}^{\oplus r}
{G_{e_1,..,e_r} \atop \longrightarrow }
{\cal O}_{l_\infty}^{\oplus r}.
\enq
Here the middle arrow is the homomorphism given by the action of ${\tilde T}$.
Using the matrixes $(B_1,B_2,I,J)$, we obtain ${\tilde T}$ action on $M(r,n)$:
\beq
(B_1,B_2,I,J) \mapsto (t_1B_1,t_2B_2,Ie^{-1},t_1t_2eJ),
\mbox{ for }t_1,t_2\in {\bf C}^*,e=\mbox{diag}(e_1,..,e_r)\in ({\bf C}^*)^r.
\enq
Here ${\tilde T}$-action preserves $\mu_{\bf C}(B_1,B_2,I,J)=0$ and the stability condition commutes the action of $GL(V)$.

In the same way, we have a ${\tilde T}$-action on $M_0(r,n)$.
The map $\pi:M(r,n)\to M_0(r,n)$ is equivalent.

The fixed points $M(r,n)^{\tilde T}$ consist of
$(E,\Phi)=(I_1,\Phi_1)\oplus\cdots\oplus(I_r,\Phi_r)$
such that
\begin{itemize}
\item[(1)]$I_\alpha$ is an ideal sheaf of 0-dimensional subscheme $Z_\alpha$ contained in ${\bf C}^2={\bf P}^2\setminus l_\infty$
(${\cal O}_{{\bf P}^2}/I_\alpha={\cal O}_{Z_\alpha}$).
\item[(2)]$\Phi_\alpha$ is an isomorphism from $(I_\alpha)_{l_\infty}$to the $\alpha$th factor of ${\cal O}_{l_\infty}^{\oplus r}$.
\item[(3)]$I_\alpha$ is fixed by the action of ${\bf C}^*\times {\bf C}^*$,
coming from that on ${\bf P}^2$.
\end{itemize}

We parametrize the fixed point set $M(r,n)^{\tilde T}$
by $r$-tuple of Young diagrams $\vec{Y}=(Y_1,..,Y_r)$.
$Y_\alpha$ corresponds to the ideal $I_\alpha$
spanned by monomials $x^{h}y^{k}$ placed at $(h+1,k+1)$
outside $Y_\alpha$.
The constraint is that the total number of boxes $|\vec{Y}|:=\sum_\alpha |Y_\alpha|$ is equal to n.

Let $Y=(\lambda_1\ge\lambda_2\ge\cdots)$ be a Young diagram,
where $\lambda_h$ is the length of the $h$-th column of $Y$,
and $\lambda_k^\prime$ is the length of the $k$-th row of $Y$.
Let $l(Y)$ denote the number of columns of $Y$,i.e.,$l(Y)=\lambda_1^\prime$.
Then, we define $l_{Y}(s)$ and $a_{Y}(s)$ as
\beq
l_Y(s):=\lambda_h-k,~~     a_Y(s):=\lambda_k^\prime-h,
\enq
where $s=(h,k)\in({\bf Z}_{>0})^2$.

On the other hand, the fixed points $M_0(r,n)^{\tilde T}$
consist of the single point $n[0]\in S^n{\bf C}^2\subset M_0(r,n)$.

\begin{thm}\cite{n-y,po}
Let $(E,\Phi)$ be a fixed point of ${\tilde T}$-action
corresponding to $\vec{Y}=(Y_1,..,Y_r)$.
Then the ${\tilde T}$-module structure of $T_{(E,\Phi)}M(r,n)$
is given by
\beq
T_{(E,\Phi)}M(r,n)=\sum_{\alpha,\beta}^{r}N_{\alpha,\beta}^{\vec{Y}}(t_1,t_2)
\enq
where
\beq
N_{\alpha,\beta}^{\vec{Y}}(t_1,t_2)=
e_\beta e_\alpha^{-1}
\times
\left\{
\sum_{s\in Y_\alpha}(t_1^{-l_{Y_\beta}(s)}t_2^{a_{Y_\alpha}(s)+1})
+\sum_{s\in Y_\beta}(t_1^{l_{Y_\alpha}(s)+1}t_2^{-a_{Y_\alpha}(s)})
\right\}.
\enq
Here we denote $e_\alpha (\alpha =1,..,r)$ by the one dimensional ${\tilde T}$-module
given by 
\[
{\tilde T}\ni (t_1,t_2,..,e_r)\mapsto e_\alpha.
\]
Similarly $t_1,t_2$ denote one-dimensional ${\tilde T}$-modules.
Then, the representation ring $R({\tilde T})$ 
is isomorphic to ${\bf Z}[t_1^{\pm},t_2^{\pm},e_1^{\pm},..,e_r^{\pm}]$,
where $e_\alpha^{-1}$ is the dual of $e_\alpha$.
\end{thm}

Now we move to the case of ${\hat M}(r,k,n)^{\tilde T}$.
We assume that $Hom(E,E(-l_\infty))=Ext^2(E,E(-l_\infty))=0$
in this article. This condition is related to  
the smoothness of its moduli space.

Let us define an action of the $(r+2)$-dimensional torus 
${\tilde T}={\bf C}^*\times{\bf C}^*\times T^r$ on ${\hat M}(r,k,n)$
by modifying the action of ${\tilde T}$ on $M(r,n)$ as follows.
For $(t_1,t_2)\in {\bf C}^*\times{\bf C}^*$,
let $F_{t_1,t_2}^\prime$ be an automorphism on $F_2$ defined by
\beq
F_{t_1,t_2}^\prime([z_0:z_1:z_2],[z:w])=([z_0:t_1^2z_1:t_2^2z_2],[t_1z:t_2w]),
\enq
where the condition $z_1w^2=z_2z^2$ is preserved.
Then we define the action of ${\tilde T}$ by replacing $F_{t_1,t_2}$ by $F_{t_1,t_2}^\prime$
in (3.25).
The action of the latter $T^r$ is exactly the same as before.
Under a pullback $p^*:F_2\to {\bf P}^2$, an automorphism on ${\bf P}^2$ is given by
\beq
p^*F_{t_1,t_2}^\prime ([z_0:z_1:z_2])=([z_0:t_1^2z_1:t_2^2z_2]).
\enq
Then, the morphism ${\hat \pi}$ is equivalent.

Note that the fixed point set of 
${\bf C}^*\times{\bf C}^*$ in $\widehat {{\bf C}^2/{\bf Z}_2}=F_2\setminus l_\infty$
consists of two points $([1:0:0],[1:0])$ and $([1:0:0],[0:1])$,
which are denoted by $p_1$ and $p_2$ respectively.

The fixed points ${\hat M}(r,n)^{\tilde T}$ consist of
$(E,\Phi)=(I_1(k_1/2),\Phi_1)\oplus\cdots\oplus(I_r(k_r/2),\Phi_r)$
such that
\begin{itemize}
\item[(1)]$I_\alpha(k_\alpha/2)$ is 
the tensor product $I_\alpha\otimes {\cal O}(k_\alpha C/2)$,
where $k_\alpha\in {\bf Z}$ and $I_\alpha$ is
an ideal sheaf of 0-dimensional subscheme $Z_\alpha$ contained 
in $\widehat{{\bf C}^2/{\bf Z}_2}=F_2\setminus l_\infty$.
\item[(2)]$\Phi_\alpha$ is an isomorphism from $(I_\alpha)_{l_\infty}$to the $\alpha$th 
factor of ${\cal O}_{l_\infty}^{\oplus r}$.
\item[(3)]$I_\alpha$ is fixed by the action of ${\bf C}^*\times {\bf C}^*$,
coming from that on $F_2$.
\end{itemize}
$Supp(Z_\alpha)=\{p_1,p_2\}\in F_2\setminus l_\infty$. 
Thus $Z_\alpha=Z_\alpha^1\cup  Z_\alpha^2$,
where $Z_\alpha^1$ and  $Z_\alpha^2$ are
supported at $p_1$ and $p_2$ respectively.
Alternatively $I_\alpha=I_\alpha^1\cap I_\alpha^2$,
where ${\cal O}/I_\alpha^k={\cal O}_{Z_\alpha^k}$.
If we take a coordinate system $(x,y)=(z_1/z_0,w/z)((z/w,z_2/z_0))$ around $p_1(p_2)$,
then $I_\alpha^1(I_\alpha^2)$ is generated by monomials $x^ky^h$.
Then $I_\alpha^1(I_\alpha^2)$ corresponds to a Young diagram $Y_\alpha^1(Y_\alpha^2)$
as before.
Therefore the fixed point set is parametrized by $r$-tuples of 
$(\vec{k},\vec{Y^1},\vec{Y^2})=((k_1,Y_1^1,Y_1^2),..,(k_r,Y_r^1,Y_r^2))$,
where $k_\alpha\in {\bf Z}$.
The constraint is
\beq
\sum_\alpha k_\alpha=k,~~
|\vec{Y^1}|+|\vec{Y^2}|
+\frac{1}{r}\sum_{\alpha<\beta}
\left|\frac{k_\alpha}{2}-\frac{k_\beta}{2}\right|^2=n.
\enq

One would wonder why we introduce fractional line bundles
${\cal O}(k_\alpha C/2)$.
${\cal O}(k_\alpha C/2)$ themselves are not defined rigorously.
However we are interested in the relation between 
affine Lie algebra and instanton calculus on $\widehat{{\bf C}^2/{\bf Z}_2}$.
Indeed ${\cal O}(-1)$ curve case \cite{n-y,n-y1,n-y2}
are related to level $1$ theta functions.
Moreover ${\cal O}(-2)$ curve case is also
related to level $2$ theta functions as mentioned in Sec.A.2.
In this case, fractional line bundles must be introduced.

Hereafter we only consider the case of $r=2$.
There are two types of $\hat{M}(2,k,n)$: 
$k \equiv 0$ mod $2$ (even type)
and $k \equiv 1$ mod $2$ (odd type).
We only consider the case of $k \equiv 0$ mod $2$ (even type).
This is because there is a difficulty in the proof of Theorem 6 in odd case. 
The proof is done in the same way as \cite{n-y1}.

\begin{thm}
Let $(E,\Phi)$ be a fixed point of  ${\tilde T}$-action
corresponding to $(\vec{k},\vec{Y}^1,\vec{Y^2})=((k_1,Y_1^1,Y_1^2),(k_r,Y_r^1,Y_r^2))$.
Suppose $k\equiv 0$ mod $2$.
Then the ${\tilde T}$-module structure of $T_{(E,\Phi)}\hat{M}(r,k,n)$
is given by
\beq
T_{(E,\Phi)}\hat{M}(r,k,n)=\sum_{\alpha,\beta}^{2}\left(
L_{\alpha,\beta}^{\vec{k}}(t_1,t_2)
+t_1^{k_\beta-k_\alpha}N_{\alpha,\beta}^{\vec{Y}^1}(t_1^2,t_2/t_1)
+t_2^{k_\beta-k_\alpha}N_{\alpha,\beta}^{\vec{Y}^2}(t_1/t_2,t_2^2)\right),
\enq
where
\beq
L_{\alpha,\beta}^{\vec{k}}(t_1,t_2)=e_\beta e_\alpha^{-1}\times 
\left\{
\begin{array}{l}
{\displaystyle
\sum_{i,j\ge 0,i+j\equiv 0 \mbox{\tiny  ~mod }2 \atop
i+j\le k_\alpha-k_\beta-2}t_1^{-i}t_2^{-j}\mbox{ if } k_\alpha>k_\beta+1,}\\
{\displaystyle\sum_{i,j\ge 0,i+j\equiv 0 \mbox{\tiny  ~mod }2 \atop
i+j\le k_\beta-k_\alpha-2}t_1^{i+1}t_2^{j+1}\mbox{ if } k_\alpha+1<k_\beta,}\\
0\mbox{ ~~~~~~~~~~~~~~~~~otherwise.
}
\end{array}
\right.
\enq
\end{thm}
{\it Proof}.  According to the decomposition 
$E=I_1(k_1C/2)\oplus\cdots I_r(k_rC/2)$,
the tangent space $T_{(E,\Phi)}\hat{M}(r,k,n)=Ext^1(E,E(-l_\infty))$
is decomposed as
\beq
Ext^1(E,E(-l_\infty))=\bigoplus_{\alpha,\beta}
Ext^1(I_\alpha(k_\alpha C/2),I_\beta(k_\beta C/2-l_\infty)).
\enq 
The factor $Ext^1(I_\alpha(k_\alpha C/2),I_\beta(k_\beta C/2-l_\infty))$
has weight $e_\beta e_\alpha^{-1}$ as a $T$-module.
We only have to describe each factor as a $T^2$-module.

Under the assumption that $Hom(E,E(-l_\infty))=Ext^2(E,E(-l_\infty))=0$,
$Ext^1(I_\alpha(k_\alpha C/2),I_\beta(k_\beta C/2-l_\infty))
=-\sum_{i=0}(-1)^i
Ext^i(I_\alpha(k_\alpha C/2),I_\beta(k_\beta C/2-l_\infty))$.
Thus, using the exact sequence
$0\to I_\alpha\to {\cal O}\to {\cal O}_{Z_\alpha}\to 0$,
we have
\beqy
&&
\sum_{i=0}(-1)^iExt^i(I_\alpha(k_\alpha C/2),I_\beta(k_\beta C/2-l_\infty))
\no
\\
&=&
\sum_{i=0}(-1)^iExt^i({\cal O}(k_\alpha C/2),{\cal O}(k_\beta C/2-l_\infty))
-\sum_{i=0}(-1)^iExt^i({\cal O}(k_\alpha C/2),{\cal O}_{Z_\beta}(k_\beta C/2-l_\infty))\no\\
&&-\sum_{i=0}(-1)^iExt^i({\cal O}_{Z_\alpha}(k_\alpha C/2),{\cal O}(k_\beta C/2-l_\infty))
+\sum_{i=0}(-1)^iExt^i({\cal O}_{Z_\alpha}(k_\alpha C/2),{\cal O}{Z_\beta}(k_\beta C/2-l_\infty)).
\no\\
\enqy

First we consider the term $\sum_{i=0}(-1)^iExt^i({\cal O}(k_\alpha C/2),{\cal O}(k_\beta C/2-l_\infty))=-Ext^1({\cal O}(k_\alpha C/2),{\cal O}(k_\beta C/2-l_\infty))=-H^1({\cal O}((k_\beta-k_\alpha)C/2-l_\infty))=-L_{\alpha,\beta}$.We set $n=k_\alpha-k_\beta$.
Since $H^1(F_2,{\cal O}(-l_\infty))=0$, we only consider two cases of $n>1$ and $n<-1$.
On the other hand, $H^1(F_2,{\cal O}(\pm C/2-l_\infty))=0$
may not happen. This is the difficulty in odd type.

Let us consider the case of $n>1$.
We consider the cohomology long exact sequence associated with
an exact sequence 
$0\to {\cal O}(-nC/2)\to {\cal O}((-n+2)C/2)\to {\cal O}_C((-n+2)C/2)\to 0$, 
which is equivalent under ${\bf C}^*\times {\bf C}^*$-action.
Since $C$ is a projective line with self-intersection $(-2)$,
we have $H^1(C,{\cal O}_C((-n+2)C/2))=H^1({\bf P}^1,{\cal O}_{{\bf P}^1}(n-2)))=0$.Thus we have
\beq
0\to H^0(C,{\cal O}_{{\bf P}^1}(n-2))\to 
H^1(F_2,{\cal O}((-n)C/2-l_\infty))\to H^1(F_2,{\cal O}((-n+2)C/2-l_\infty))\to 0,
\enq
which is an exact sequence in ${\bf C}^*\times {\bf C}^*$-modules.
Starting with $H^1(F_2,{\cal O}(-l_\infty))=0$, we obtain
\beq
H^1(F_2,{\cal O}(-nC/2-l_\infty))=\bigoplus_{d=0}^{n/2-1}H^0({\bf P}^1,{\cal O}_{{\bf P}^1}(2d))
\enq
by induction. Here $n$ is  even in this case. 
Since $H^0({\bf P}^1,{\cal O}_{{\bf P}^1}(d))$ is the space 
of homogeneous polynomials in $z,w$ of degree $d$,
it is equal to $\sum_{i=0}^{d}t_1^{-i}t_2^{-d+i}$
in the representation ring  of $T^2$. Thus we have
\beq
L_{\alpha,\beta}^{n>1}=\sum_{d=0}^{n/2-1}\sum_{i=0}^{2d}t_1^{-i}t_2^{-2d+i}=\sum_{i,j\ge 0,i+j\equiv 0 \mbox{\tiny ~mod }2 \atop i+j\le n-2}t_1^{-i}t_2^{-j}.
\enq

Next we consider the case of $n<-1$.
We use 
$0\to {\cal O}((-n-2)C/2)\to {\cal O}(-nC/2)\to {\cal O}_C(-nC/2)\to 0$
to obtain
\beq
0\to H^1(F_2,{\cal O}((-n-2)C/2-l_\infty))\to
H^1(F_2,{\cal O}(-nC/2-l_\infty)) \to
H^1({\bf P}^1,{\cal O}_{{\bf P}^1}(n))\to 0.
\enq 
Here we use $H^0({\bf P}^1,{\cal O}_{{\bf P}^1}(n))=0$ in this case.
Starting with $H^1(F_2,{\cal O}((-n-2)C/2-l_\infty))$ for $n=-2$,
we obtain 
\beq
H^1(F_2,{\cal O}(-nC/2-l_\infty))=\bigoplus_{d=1}^{-n/2}H^1({\bf P}^1,{\cal O}_{{\bf P}^1}(-2d))
\enq
by induction.
To use the Serre duality for $H^1({\bf P}^1,{\cal O}_{{\bf P}^1}(-2d))$,
we give the canonical bundle $K_{{\bf P}^1}\cong t_1^{-1}t_2^{-1}{\cal O}_{{\bf P}^1}(-2)$. Using this, $H^1({\bf P}^1,{\cal O}_{{\bf P}^1}(-2d))$
is the dual of  $t_1^{-1}t_2^{-1}H^0({\bf P}^1,{\cal O}_{{\bf P}^1}(2d-2))$.
Thus we have
\beq
L_{\alpha,\beta}^{n<-1}=t_1t_2\sum_{d=1}^{-n/2}\sum_{i=0}^{2d-2}t_1^{i}t_2^{2d-2-i}=\sum_{i,j\ge 0,i+j\equiv 0 \mbox{\tiny ~mod }2 \atop i+j\le -n-2}t_1^{i+1}t_2^{j+1}.
\enq

Now we move to the remaining three terms in (3.39).
We have

\beqy
&&
-\sum_{i=0}(-1)^iExt^i({\cal O}(k_\alpha C/2),{\cal O}_{Z_\beta}(k_\beta C/2-l_\infty))-\sum_{i=0}(-1)^iExt^i({\cal O}_{Z_\alpha}(k_\alpha C/2),{\cal O}(k_\beta C/2-l_\infty))\no\\
&&
+\sum_{i=0}(-1)^iExt^i({\cal O}_{Z_\alpha}(k_\alpha C/2),{\cal O}_{Z_\beta}(k_\beta C/2-l_\infty))
\no\\
&=&
-\sum_{i=0}(-1)^iExt^i({\cal O},{\cal O}_{Z_\beta}((k_\beta -k_\alpha)C/2-l_\infty))-\sum_{i=0}(-1)^iExt^i({\cal O}_{Z_\alpha}((k_\alpha-k_\beta) C/2),{\cal O}(-l_\infty))\no\\
&&
+\sum_{i=0}(-1)^iExt^i({\cal O}_{Z_\alpha},{\cal O}_{Z_\beta}((k_\beta-k_\alpha) C/2-l_\infty)).
\enqy
Since we have a decomposition $Z_\alpha=Z_\alpha^1 \cup Z_\alpha^2$,
we obtain the remaining terms in (3.46) as the direct sum of 
the corresponding terms for $Z_\alpha^1(Z_\beta^1)$ and $Z_\alpha^2(Z_\beta^2)$.
This is because mixed terms such as 
$\sum_{i=0}Ext^i({\cal O}_{Z_\alpha^1}(k_\alpha C/2),{\cal O}_{Z_\beta^2}(k_\beta C/2-l_\infty))$ are zero.

First we consider the terms for $Z_\alpha^1(Z_\beta^1)$.
We take a coordinate system $(x,y)=(z_1/z_0,w/z)$,
which transforms $(t_1^2x,t_2/t_1y)$ under $T^2$-action.
Since the divisor $C$ is given by $x=0$,
the multiplication by $x^m$ induces an isomorphism ${\cal O}_{Z_\alpha^1}(mC)\cong {\cal O}_{Z_\alpha^1}$ of sheaves for $m\in {\bf Z}/2$.
For an isomorphism of equivalent sheaves, we twist it as 
${\cal O}_{Z_\alpha^1}(mC)\cong t_1^{2m}{\cal O}_{Z_\alpha^1}$.
Hence we sum up (3.46) for $p_1$ and obtain
\beq
t_1^{k_\beta-k_\alpha}\left(-\sum_{i=0}(-1)^iExt^i({\cal O},{\cal O}_{Z_\beta^1}(-l_\infty))
-\sum_{i=0}(-1)^iExt^i({\cal O}_{Z_\alpha^1},{\cal O}(-l_\infty))
+\sum_{i=0}(-1)^iExt^i({\cal O}_{Z_\alpha^1},{\cal O}_{Z_\beta^1}(-l_\infty))
\right)
\enq
Since $Z_\alpha^1$ is supported at the single point $p_1$,
we can consider it as a subscheme of ${\bf P}^2$ 
supported at the origin $[1:0:0]$,
where $T^2$-action on ${\bf P}^2$ is
$[z_0:z_1:z_2]\mapsto [z_0:t_1^2z_1:t_2/t_1z_2]$.
Let $I_\alpha^1$ be the corresponding ideal sheaves of ${\cal O}_{{\bf P}^2}$.
Using the $0\to I_\alpha^1\to {\cal O}_{{\bf P}^2} \to {\cal O}_{{\bf P}^2}/I_\alpha^1={\cal O}_{Z_\alpha^1}\to 0$, we obtain
\beq
t_1^{k_\beta-k_\alpha}\left(
\sum_{i=0}(-1)^iExt^i(I_\alpha^1,I_\beta^1)
-\sum_{i=0}(-1)^iExt^i({\cal O}_{{\bf P}^2},{\cal O}_{{\bf P}^2}(-l_\infty))
\right).
\enq
The second term $\sum_{i=0}(-1)^iExt^i({\cal O}_{{\bf P}^2},{\cal O}_{{\bf P}^2}(-l_\infty))$ is zero.
Thus we can use Theorem 5 after replacing $(t_1,t_2)$
by $(t_1^2,t_2/t_1)$,
and obtain $N_{\alpha}^{\vec{Y^1}}(t_1^2,t_2/t_1)$.

The terms for $Z_\alpha^2(Z_\beta^2)$ can be calculated in the similar way. 
$\rotatebox[origin=r]{45}{\huge $\diamond$}$

The naive estimation of the tangent space $T_{(E,\Phi)}\hat{M}(2,k,n)$
of $\hat{M}(2,k,n)$ for odd $k$ is done.
The result is given in Sec.C.
In this estimation, we neglect the obstruction
coming from $H^1(F_2,{\cal O}(\pm C/2-l_\infty))\ne 0$.
We have to estimate the tangent space $T_{(E,\Phi)}\hat{M}(2,2m+1,n)$ precisely
by considering the contribution from  $H^1(F_2,{\cal O}(\pm C/2-l_\infty))\ne 0$.After estimating the tangent space $T_{(E,\Phi)}\hat{M}(2,2m+1,n)$,
we can derive  Poincar\'e polynomials of $\hat{M}(2,2m+1,n)$.
Furthermore, since the tangent space $T_{(E,\Phi)}\hat{M}(r,k,n)$
is decomposed as (3.38),
we can estimate the tangent space $T_{(E,\Phi)}\hat{M}(r,k,n)$
for general $r,k$,
by using the data of $T_{(E,\Phi)}\hat{M}(2,2m,n)$ and $T_{(E,\Phi)}\hat{M}(2,2m+1,n)$.
Therefore we can also derive  Poincar\'e polynomials of $\hat{M}(r,k,n)$
for general $r,k$ straightforwardly.

\subsection{Betti Numbers of ${\hat M}(2,2m,n)$}
Using the results of the previous subsection,
we can calculate Betti numbers of $M(r,n)({\hat M}(2,2m,n))$.
Furthermore we obtain the generating function of them 
as a beautiful closed  formula.
An algorithm of calculation of 
Betti numbers of $M(r,n)$
is as follows\cite{n-y}.
For this purpose, we introduce one parameter subgroup 
$\lambda:{\bf C}^*\to {\tilde T}$ as
\beq
\lambda(t)=(t^{m_1},t^{m_2},t^{n_1},..,t^{n_r}).
\enq
If we choose weights $m_1,m_2,n_\alpha$ generic,
the Zariski closure of $\lambda({\bf C}^*)$ is equal to the whole ${\tilde T}$,
and the fixed point set of $\lambda({\bf C}^*)$ coincides with that of ${\tilde T}$.
For weights $m_1,m_2,n_\alpha$, which satisfy
\beq
m_2>>n_1>n_2>\cdots >n_r>>m_1>0,
\enq
the fixed points satisfy
$M(r,n)^{\lambda({\bf C}^*)}=M(r,n)^{\tilde T}$.
We denote a fixed point by $(E,\Phi)\in M(r,n)^{\tilde T}$.
The tangent space at the fixed point 
$T_{(E,\Phi)}M(r,n)$
has a ${\tilde T}$-module structure 
and an induced ${\bf C}^*$-structure 
via $\lambda$ satisfying (3.50). 
Using Theorem 5,
the index $D$ of the fixed point is given by twice 
the sum of the numbers of 
$s\in Y_{\alpha,\beta}$ for which the weight for ${\bf C}^*$ is negative
\cite{nak}.
Considering the fixed points satisfying the constraint $|\vec{Y}|=n$,
we obtain Poincar\'e polynomial $P_z(M(r,n))$ of $M(r,n)$
 by summing up $z^{D}$.
Poincar\'e polynomial of $\hat{M}(2,2m,n)$ is also obtained 
in the similar way.
However we consider another algorithm of calculation 
of Betti numbers of $\hat{M}(2,2m,n)$
by choosing a special case\cite{n-y}:
\beq
m_1=m_2>>n_1>n_2>0,
\enq
where $m_1,n_\alpha$ are generic.
Since $\lambda$ is not generic,
the fixed points are different from those of ${\tilde T}$.
We also denote these points by $(E,\Phi)=(I_1(k_1C/2),\Phi_1)\oplus(I_2(k_2C/2),\Phi_2) $,
but they satisfy the following conditions:
\begin{itemize}
\item[(1)]$I_\alpha(k_\alpha/2)$ is 
the tensor product $I_\alpha\otimes {\cal O}(k_\alpha C/2)$,
where $k_\alpha\in {\bf Z}$ and $I_\alpha$ is
an ideal sheaf of 0-dimensional subscheme $Z_\alpha$ contained 
in $\widehat{{\bf C}^2/{\bf Z}_2}=F_2\setminus l_\infty$.
\item[(2)]$\Phi_\alpha$ is an isomorphism from $(I_\alpha)_{l_\infty}$to the $\alpha$th 
factor of ${\cal O}_{l_\infty}^{\oplus r}$.
\item[(3)]$I_\alpha$ is fixed by 
the diagonal subgroup $\Delta {\bf C}^*$ of ${\bf C}^*\times {\bf C}^*$,
coming from that on $F_2$.
\end{itemize}
The difference between (3.50) and (3.51)
appears in these fixed points,
but the sum of each contribution is the same. 
A typical difference can be seen in rank one case.
The Hilbert scheme of point on surface appears in (3.50) case,
while the Hilbert scheme of point on cotangent bundle of Riemann surface appear 
in (3.51) case\cite{nak}.
The components of the fixed point set are parametrized
by $(\vec{k},\vec{Y})=((k_1,Y_1),(k_2,Y_2))$,
whose constraint is 
\beq
\sum_\alpha k_\alpha=k,~~ |\vec{Y}|+\frac{1}{r}
\left|\frac{k_1}{2}-\frac{k_2}{2}\right|^2=n.
\enq
Note that $\vec{Y^1}$ in (3.35) are set to be zero
due to the property of $I_\alpha$.
Now we see the property of $I_\alpha$ in detail.
A general point in the component $(\vec{k},\vec{Y})$ is
$(E,\Phi)=(I_1(k_1C/2),\Phi_1)\oplus (I_2(k_2C/2),\Phi_2)$
such that
\begin{itemize}
\item[(1)]the support of $I_\alpha$ consists of 
$P_1,P_2,..,P_{l(Y_\alpha)}$, contained in the exceptional curve $C$,
\item[(2)] if $\xi$ is the inhomogeneous coordinate of $C={\bf P}^1$
and $\eta$ is the coordinate of the fiber 
$\widehat{{\bf C}^2/{\bf Z}_2}\cong {\cal O}(-2)\to C$,
\beq
I_\alpha=(\xi-\xi_1,\eta^{\lambda_1^\alpha})\cap
(\xi-\xi_2,\eta^{\lambda_2^\alpha})\cap \cdots \cap 
(\xi-\xi_{l(Y_\alpha)},\eta^{\lambda_{l(Y_\alpha)}^\alpha}),
\enq
where $x_l=\xi(P_l)$.
\end{itemize}
The points $P_1,P_2,..,P_{l(Y_\alpha)}$ move in ${\bf P}^1$,
but their order is irrelevant when the values ${\lambda_l^\alpha}$ are the same. Thus the component is isomorphic to 
\beq
S^{Y_1}{\bf P}^1\times \cdots \times S^{Y_r}{\bf P}^r.
\enq
We define $S^{Y}{\bf P}^1$ in the following way.
For a Young diagram $Y=(\lambda_1\ge \lambda_2\ge ...)$,
we denote 
\beq
Y=(1^{m_1}2^{m_2}\cdots), 
\enq
where $m_i=\#\{l| \lambda_l=i \}$. Using this notation, $S^{Y}{\bf P}^1$ is 
given by
\beq
S^{Y}{\bf P}^1=S^{m_1}{\bf P}^1\times S^{m_1}{\bf P}^1\times \cdots
={\bf P}^{m_1}\times {\bf P}^{m_2}\times\cdots.
\enq
Note that $S^{m}{\bf P}^1={\bf P}^{m}$.

Suppose $k\equiv 0$ mod 2.
Let $(E,\Phi)$ be a fixed point in the component
corresponding to $((k_1,Y_1),(k_2,Y_2))$.
Then, the tangent space $T_{(E,\Phi)}\hat{M}(2,2m,n)$
is a $\Delta {\bf C}^*\times T^r$-module.
Using Theorem 6, we obtain 
\beq
T_{(E,\Phi)}\hat{M}(2,2m,n)=\sum_{\alpha,\beta}
(L_{\alpha,\beta}^{\vec{k}}(t_1,t_1)
+t^{k_\beta-k_\alpha}N_{\alpha,\beta}^{\vec{Y}}(1,t_1^2)).
\enq
Here we set $\vec{Y^1}=0$. Using this, we obtain
\begin{thm}
{\it 
The poincar\'e polynomial of ${\hat M}(2,2m,n)$ is given by}
\begin{equation}
P_t({\hat M}(2,2m,n))=\sum_{|\vec{Y}|+\frac{1}{2}\left|\frac{k_1}{2}-\frac{k_2}{2}\right|^2=n, \atop \sum_\alpha k_\alpha=2m}
 \prod_{\alpha=1}^2 t^{2(|Y_\alpha|-l(Y_\alpha))}
P_t(S^{Y_\alpha}{\bf P}^1)
\cdot t^{2(l^\prime+|Y_\alpha|+|Y_\beta|-n^\prime)},
\end{equation}
where 
\begin{equation}
l^\prime=\left\{
\begin{array}{c}
(K+m)^2 \mbox{ if }k_\alpha=K+2m\ge k_\beta=-K,\\
(-K-m)^2-1 \mbox{ if } k_\alpha=K+2m < k_\beta=-K,
\end{array}
\right.
\end{equation}
\begin{equation}
n^\prime=\left\{
\begin{array}{c}
(\# \mbox{ of columns of $Y_\alpha$ which are longer than $K+m)$ if $K+m\ge 0$},\\
(\#  \mbox{ of columns of $Y_\beta$ which are longer than $-K-m-1)$ if $K+m < 0$}.
\end{array}
\right.
\end{equation}
\end{thm}
The calculation for rank one is done in the same way.
This implies that $P_t(\hat{M}(1,0,n))=P_t((\widehat{{\bf C}^2/{\bf Z}_2})^{[n]})$.
On the other hand, under the action of (3.34) the fixed points $M_0(1,n)^{\tilde T}$
consist of the single point $n[0]\in S^n{\bf C}^2\subset M_0(1,n)$.
Thus, there is a projective morphism:
\beq
\hat{M}(1,0,n)=(\widehat{{\bf C}^2/{\bf Z}_2})^{[n]}\to S^n{\bf C}^2=M_0(1,n).
\enq
This is an evidence of Conjecture 1.

Concretely we give the generating function of the case of $m=0$.\\
{\bf Theorem 1} {\it
The generating function of Poincar\'e polynomial is}
\begin{eqnarray}
&&\sum_{n\in \frac{{\bf Z}_{\ge 0}}{2}}P_t({\hat M}(2,0,n))q^n
\nonumber \\
=&&
\prod_{d=1}^\infty\frac{1}{(1-t^{4d}q^d)(1-t^{4d-2}q^d)^2(1-t^{4d-4}q^d)}
\nonumber \\
&& \times [\sum_{k\ge 0}\prod_{d=1}^k\frac{1-t^{4d-4}q^d}{1-t^{4d}q^d}
t^{2k^2}q^{\frac{k^2}{2}}
+\sum_{k > 0}\prod_{d=1}^{k-1}\frac{1-t^{4d-4}q^d}{1-t^{4d}q^d}
t^{2k^2-2}q^{\frac{k^2}{2}}
]
\\
=&&
\prod_{d=1}^\infty\frac{1}{(1-t^{4d}q^d)(1-t^{4d-2}q^d)^2(1-t^{4d-4}q^d)}
\prod_{d=1}^\infty \frac{1-(-t^2q^{\frac{1}{2}})^dt^{-2}}{1-(-t^2q^{\frac{1}{2}})^d}
\sum_{k\in{\bf Z}}t^{2k^2}q^{\frac{k^2}{2}}
\no\\
&&
\\
=&&
\prod_{d=1}^\infty\frac{1}{(1-t^{4d}q^d)(1-t^{4d-2}q^d)^2(1-t^{4d-4}q^d)}
\prod_{d=1}^\infty \frac{1-(-t^2q^{\frac{1}{2}})^dt^{-2}}{1+(-t^2q^{\frac{1}{2}})^d}.
\end{eqnarray}
The equality between (3.62) and (3.63) is verified in Sec.B.
The middle factor in (3.63) is a typical difference from
the twisted ${\cal N}=4$ partition function.
This factor causes the failure of describing the generating function
by using level two affine Lie algebras\cite{kac}(See also Sec.A).
The second factor in (3.64)
is corresponding to the contribution from vector bundles.
We also calculate the generating function of Poincar\'e polynomial
in the different way in the next section.
There is also an identity.

\begin{lem}
\begin{equation}
\sum_{n\in \frac{{\bf Z}_{\ge 0}}{2}}P_t({\hat M}(2,2m,n))q^n=
\sum_{n\in \frac{{\bf Z}_{\ge 0}}{2}}P_t({\hat M}(2,0,n))q^n.
\end{equation}
\end{lem}

To obtain the full $U(2)$ generating function, we define
\beq
Z_{A_1}^{U(2)}(q,z):=\sum_{m,n\in {\bf Z}}P_t({\hat M}(2,2m,n))q^{n+\frac{(2m)^2}{8}}z^{2m}.
\enq
By using Lemma 3 and Sec.A, this formula can be rewritten by
\begin{eqnarray}
Z_{A_1}^{U(2)}(q;z)&=&\sum_{m,n\in {\bf Z}}P_t({\hat M}(2,2m,n))q^{n+\frac{(2m)^2}{8}}z^{2m}
\nonumber
\\
&=&
\sum_mq^{\frac{m^2}{2}}z^{2m}\sum_{n\in \frac{{{\bf Z}\ge 0}}{2}}P_t({\hat M}(2,0,n))q^{n}
\no\\
&=&
\theta_3(q;z)\sum_{n\in \frac{{{\bf Z}\ge 0}}{2}}P_t({\hat M}(2,0,n))q^{n}
\end{eqnarray}
On the other hand, the $U(1)$ generating function of Poincar\'e polynomial
on $\widehat{{\bf C}^2/{\bf Z}_2}$ 
was already calculated in \cite{Fucito,Fujii}.
Following the method used in \cite{Fucito,Fujii}, 
we calculate the first several part of
the $U(2)$ generating function of Poincar\'e polynomial
and obtain the same result.
Recently the closed formula of the $U(2)$ generating function 
of Poincar\'e polynomial
was proposed by T.Hausel(See (9) in\cite{hausel}). 
His result also coincides with the above result.

Finally we consider the parameters which satisfy
\beq
n_1>>n_2>>m_2>>m_1>0.
\enq
As mentioned in \cite{Fucito}, this situation corresponds to
the $U(1)^2$ case instead of the $U(2)$ case.
The former corresponds to the Coulomb branch,
while the latter corresponds to the Higgs branch.
We denote the framed moduli space for $U(1)^2$ by $\hat{M}^\prime(2,2m,n)$
satisfying (3.35).Using Theorem 6, we obtain
\begin{thm}
\beq
P_t(\hat{M}^\prime(2,0,n))=\sum_{|\vec{Y}|+\frac{1}{2}\left|\frac{k_1}{2}-\frac{k_2}{2}
\right|^2=n,\atop \sum_\alpha k_\alpha=0}
\prod_{\alpha=1}^2t^{2(r|Y_\alpha^1|-l(Y_\alpha^1)+r|Y_\alpha^2|)}
\cdot t^{2(\frac{k_\alpha-k_\beta}{2})^2}.
\enq
\end{thm}
Furthermore we obtain
\begin{lem}
\beq
\sum_{n\in \frac{\bf Z}{2}}P_t(\hat{M}^\prime(2,0,n))q^n
=\frac{\theta_3(t^4q;1)}{\prod_{d=1}(1-t^{4d-2}q^d)^2(1-t^{4d}q^d)^2}.
\enq
\end{lem}
This formula coincides with (3.64) under $t=-1$,
which implies that the both formula is the same in the partition function level. We remark that the partition function is described
as a level two $A_1$ theta function over eta functions up to $q$ power.  
Following the full $U(2)$ partition function,
we  multiplying (3.70) with $t=-1$, by $\theta_3(q;1)$
to obtain the same full $U(2)$ partition function as before.
The resulting formula implies 
that the $U(2)$ partition function on $\widehat{{\bf C}^2/{\bf Z}_2}$
is the second power of the $U(1)$ partition function
on $\widehat{{\bf C}^2/{\bf Z}_2}$.
The first factor of the full $U(2)$ partition function
comes from the untwisted sector,
and the second factor comes from the twisted sector.
Each factor is the same.
This is a typical structure of the partition function on $\widehat{{\bf C}^2/{\bf Z}_2}$.
The partition function on $\widehat{{\bf C}^2/{\bf Z}_m}$
must not have this structure in general.

\section{Betti Numbers of Moduli Spaces of
Rank Two Stable Sheaves on $\widehat{X}$}
\setcounter{equation}{0}
We give a universal relation
between the generating function of Poincar\'e
polynomial of the moduli spaces
of rank two stable sheaves
on $X$ with an $A_1$ singularity at $p\in X$
and that on $\hat{X}$ blow-uped at $p$.
We call this relation the ${\cal O}(-2)$ blow-up formula.
In the derivation, we have a difficulty in the precise definition
of stable vector bundles(sheaves) on a singular surface.
Instead of discussing this difficult point,
we consider the formal generalization of the ${\cal O}(-1)$ case
\cite{yoshioka} to the ${\cal O}(-2)$ case,
and check this ${\cal O}(-2)$ blow-up formula
in the concrete case of $X={\bf C}^2/{\bf Z}_2$,
by comparing the result in the previous section.
Here, the moduli space of
stable vector bundles(sheaves) itself on ${\bf C}^2/{\bf Z}_2$
is derived by considering ${\bf Z}_2$-invariant part of 
the moduli space of stable vector bundles(sheaves) on ${\bf C}^2$.
Main tools used in this section
are Weil conjecture and elementary transformations. 
Thus all calculations are done over finite field.
We separate our calculation into two parts:
vector bundle part and non-vector bundle part.
Since a stable sheaf $E$ has a stable vector bundle $E^{\vee\vee}$
and $E$ is naturally embedded into  $E^{\vee\vee}$,
there is an exact sequence
\beq
0\to E \to  E^{\vee\vee} \to E^{\vee\vee}/E \to 0.
\enq
Thus we count the numbers of stable vector bundle $E^{\vee\vee}$
and those of maps $E^{\vee\vee} \to E^{\vee\vee}/E$,
which correspond to vector bundle part and non-vector bundle part
respectively. 

\subsection{Moduli Space of Stable Vector Bundles}
To consider the moduli space of stable vector bundles,
let us remember stable torsion free sheaves $E$ on ${\bf P}^2$\cite{oko}.
First we define $\mu(E)(\mbox{resp. }p_E(k))$ 
in order to define a stable torsion free sheaf 
in the sense of Mumford and Takemoto\cite{Mum,Take}
(resp. in the sense of Gieseker and Maruyama\cite{Gie,Maru}):
$\mu(E):=c_1(E)/rk(E)(\mbox{rep. }p_E(k):=\chi(E\otimes{\cal O}_{{\bf P}^2}(k))/rk(E))$. 
We call $E$ a stable torsion free sheaf 
in the sense of Mumford and Takemoto
(in the sense of Gieseker and Maruyama)
if $\mu(F)<\mu(E)(p_F(k)<p_E(k)$ for $k>>0)$
for all coherent subsheaves $F\subset E$
with $0<rk F < rk E$.
We also call $E$ a semistable torsion free sheaf
if $\mu(F)\le\mu(E)(p_F(k)\le p_E(k)$ for $k>>0)$
for all coherent subsheaves $F\subset E$
with $0<rk F < rk E$.
To distinguish a stable sheaf
in the sense of Gieseker from that in the sense of Mumford,
we call the latter $\mu$-stable sheaf hereafter. 

Let $X$  be a surface over ${\bf F}_q$
with an $A_1$-singularity at $p\in X$
and $H$ be an ample divisor on $X$.
Suppose that a divisor $L$ on $X$ satisfies $(L,H)=$ odd.
We consider one point blowing-up of $X$ at $p\in X$:
$\phi:{\hat X} \to X$.
Let $C$ be the exceptional divisor with $C^2=-2$.
Then we pull back $L$ by $\phi^*$,
\beq
\begin{array}{ccccc}
\phi^*:&Pic(X)&\to&Pic({\hat X})&\\
                                                                                &\rotatebox[origin=c]{90}{$\in $}&       &\rotatebox[origin=c]{90}{$\in $}\\
                                                                                                                                   &L           &{\sim\atop\to}  &\phi^*L.
\end{array}
\enq
Using the equivalence $L\cong \phi^* L$,
we denote $\phi^* L$ by $L$ again.
Let $E$ be  a rank two vector bundle on ${\hat X}$ with
$c_1(E)=L+aC,(a=0,-1/2,-1,-3/2), c_2(E)=n$.

If $E$ is $\mu$-stable with respect to $lH-C$ 
for sufficiently large $l\in {\bf Z}_{>0}$,
we introduce these
moduli spaces $M_{lH-C}(L+aC,n)_0=:M_{H_\infty}(L+aC,n)_0$.
Similarly we introduce the moduli space of $\mu$-stable vector bundle
with respect to $H$ as $M_{H}(L+aC,n)_0$.
We assume that $M_{H_\infty}(L+aC,n)_0$
and $M_{H}(L+aC,n)_0$ are compatible
and deeply connected
in the same way as the case of $X$= a smooth four surface 
and $\hat{X}$= the blow-up of $X$\cite{Br}.

\begin{defi}
${\tilde M}_{i,n}(d)$ is the set of vector bundles $E$ such that
\beq
E|_{C}={\cal O}_C(d+i)\oplus {\cal O}_C(-d)
\enq
and $E\in M_{H_\infty}(L-iC,n)_0({\bf F}_q)$.
Then $\# M_{H}(L-iC/2,n)_0=\#{\tilde M}_{i,n}(0),i=0,1,\# M_{H}(L-iC/2,n)_0=\#{\tilde M}_{i,n}(-1),i=2,3$.
\end{defi}

First we consider the case of $a=0,2$.
For an element $E$ of $M_{H_\infty}(L,n)_0({\bf F}_q)$
and a subjection $\varphi_d:E\to {\cal O}_C(d)$ with $d\in {\bf Z}$,
we set $E^\prime = ker(\varphi_d)$.
$E^\prime$ is an elementary transformation of $E$
along ${\cal O}_C(d)$. $E$ and $E^\prime$ are related by
the following commutative diagram:
\beq
\begin{array}{ccccccccc}
 & &0& &0& & & &\\
 & &\uparrow& &\uparrow& \varphi_{d|C}&& &\\
 0&\to&{\cal O}_C(-d)&\to&E_{|C}&\to&{\cal O}_C(d)&\to&0\\
  &   &~~~~~~\uparrow \varphi_{-d}& & \uparrow &\varphi_{d}&\rotatebox[origin=c]{90}{$= $}&&\\
 0&\to&E^\prime&\to&E&\to&{\cal O}_C(d)&\to &0\\
 &&\uparrow& &\uparrow&&\\
 &&E(-C)&=&E(-C)&\\
 &&\uparrow&&\uparrow&\\
 &&0&&0\\
\end{array}
\enq
We can easily obtain
\beqy
c_1(E^\prime)&=&c_1(E)-C=L-C,\\
c_2(E^\prime)&=&c_2(E)+d=n+d.
\enqy
We remark that the correspondence between $(E,\varphi_d)$
and $(E^\prime,\varphi_{-d})$ is bijective.

To count vector bundles, we define the following set of maps:
\begin{defi}
\beq
{\tilde \Phi}_d(E)=\{ \varphi| \varphi \mbox{  is a surjection } E \to {\cal O}_C(d)\
 \}/\mbox{Aut}(E).
\enq
\end{defi}
\begin{defi}
For $F={\cal O}_C(d+c)\oplus {\cal O}_C(-d)$ with $c=0,1,2,3$,
\beq
\Phi_d(F)=\{ \varphi| \varphi \mbox{  is a surjection } F \to {\cal O}_C(d) \}/{\bf \
F}_q^X,
\enq
where for $c=0,1,$ $d\ge 0$ and for $c=2,3,$ $d\ge -1$.
 \end{defi}
For the set, we have
\begin{lem}
For $c=0$,
\beq
\# \Phi_a(F)=
\left \{
\begin{array}{cc}
q+1,&\mbox{ if } d=a=0,\\
1,&\mbox{ if } d>0,a=-d,\\
q^{2a+1},&\mbox{ if } d>0,a=d,\\
q^{2a-1}(q^2-1),&\mbox{ if } d\ge 0,a>d,\\
0, & \mbox{ otherwise}.
\end{array}
\right.
\enq
For $c=1$,
\beq
\# \Phi_a(F)=
\left \{
\begin{array}{cc}
1,&\mbox{ if } d\ge 0,a=-d,\\
q^{2a},&\mbox{ if } a=d+1\ge 1,\\
q^{2a-2}(q^2-1),&\mbox{ if } a> d+1\ge 1,\\
0, & \mbox{ otherwise}.
\end{array}
\right.
\enq
For $c=2$,
\beq
\# \Phi_a(F)=
\left \{
\begin{array}{cc}
1,&\mbox{ if } d\ge 0,a=-d,\\
q^{2a-1},&\mbox{ if } a=d+2\ge 2,\\
q^{2a-3}(q^2-1),&\mbox{ if } a> d+2\ge 1,\\
q+1,&a=-d=1,\\
0, & \mbox{ otherwise}.
\end{array}
\right.
\enq
For $c=3$,
\beq
\# \Phi_a(F)=
\left \{
\begin{array}{cc}
1,&\mbox{ if } d\ge -1,a=-d,\\
q^{2a-2},&\mbox{ if } a=d+3\ge 2,\\
q^{2a-4}(q^2-1),&\mbox{ if } a> d+3\ge 2,\\
0, & \mbox{ otherwise}.
\end{array}
\right.
\enq
\end{lem}
For $c=0,1$, Lemma 6 was already given in Lemma 1.5. in \cite{yoshioka}.

Since Aut$(E)={\bf F}_q^X$, ${\tilde \Phi}_d(E)=\Phi_d(E|_C)$.
By the above diagram, we obtain
\beq
\sum_{E\in M_{H_\infty}(L-C,n)}\#{\tilde \Phi }_{-d}(E)
=\sum_{E\in M_{H_\infty}(L,n-d)}\#{\tilde \Phi }_{d}(E),(d\ge 0),
\enq
\beq
\sum_{E\in M_{H_\infty}(L,n)}\#{\tilde \Phi }_{-d}(E)
=\sum_{E\in M_{H_\infty}(L-C,n-d)}\#{\tilde \Phi }_{d}(E),(d\ge 1).
\enq
Using $M_{H_\infty}(L,n)_0({\bf F}_q)=\coprod_{l=0}^{n}{\tilde M}_{0,n}(l)$ 
and $M_{H_\infty}(L-C,n)_0({\bf F}_q)=\coprod_{l=-1}^{n}{\tilde M}_{2,n}(l)$,
 we rewrite the above relations as follows:
\begin{lem}
\beq
\# {\tilde M}_{0,n}(d)
=\sum_{l=-1}^{d-3}q^{2d-3}(q^2-1)\# {\tilde M}_{2,n-d}(l)
+q^{2d-1}\# {\tilde M}_{2,n-d}(d-2),(d\ge 2),
\enq
\beq
\# {\tilde M}_{2,n}(d)
=\sum_{l=0}^{d-1}q^{2d-1}(q^2-1)\# {\tilde M}_{0,n-d}(l)
+q^{2d+1}\# {\tilde M}_{0,n-d}(d),(d\ge 1),
\enq
\beq
\# {\tilde M}_{0,n}(1)
=(q+1)\# {\tilde M}_{2,n-1}(-1),
\enq
\beq
\# {\tilde M}_{2,n}(0)
=(q+1)\# {\tilde M}_{0,n}(0).
\enq
\end{lem}
By using these lemmas, we find
\beq
\# M_{H_\infty}(L,n)_0({\bf F}_q)
=\sum_{k=0}^nB_k^0(q)\# {\tilde M}_{0,n-k}(0)
+\sum_{k=0}^nB_k^2(q)\# {\tilde M}_{2,n-k-1}(-1),
\enq
\beq
\# M_{H_\infty}(L-C,n)_0({\bf F}_q)
=\sum_{k=0}^nB_k^2(q)\# {\tilde M}_{0,n-k}(0)
+\sum_{k=0}^nB_k^0(q)\# {\tilde M}_{2,n-k}(-1).
\enq
We can arrange these formulas in the following form
\beqy
&&\sum_n(\# M_{H_\infty}(L,n)_0({\bf F}_q)
+t^{1/2}\# M_{H_\infty}(L-C,n)_0({\bf F}_q))t^n
\no
\\
=&&(\sum_n (B_n^0(q)+t^{1/2}B_n^2(q))t^n)
(\sum_n(\# M_H(L,n)_0({\bf F}_q)+t^{1/2}\# M_H(L-C,n)_0({\bf F}_q))t^n).
\no\\&&
\enqy
$\sum_n (B_n^0(q)+t^{1/2}B_n^2(q))t^n$ itself is given by
\begin{thm}
\beq
\sum_n (B_n^0(q)+t^{1/2}B_n^2(q))t^n
=\prod_{d\ge 1}\frac{1-q^{d-1}(-t^{1/2})^{d}}{1+q^{d}(-t^{1/2})^{d}}.
\enq
\end{thm}

Next we consider the case of $a=1,3$.
For an element $E$ of $M_{H_\infty}(L-C/2,n)_0({\bf F}_q)$
and a surjection
$\varphi_{d+1}:E\to {\cal O}_C(d+1)$ with $d\in {\bf Z}$,
we set $E^\prime = ker(\varphi_{d+1})$.
$E^\prime$ is an elementary transformation of $E$
along ${\cal O}_C(d+1)$. $E$ and $E^\prime$ are related by
the following commutative diagram:
\beq
\begin{array}{ccccccccc}
 & &0& &0& & & &\\
 & &\uparrow& &\uparrow& \varphi_{d+1|C}&& &\\
 0&\to&{\cal O}_C(-d)&\to&E_{|C}&\to&{\cal O}_C(d+1)&\to&0\\
  &   &~~~~~~\uparrow \varphi_{-d}& & \uparrow &\varphi_{d+1}&\rotatebox[origin=c]{90}{$= $}&&\\
 0&\to&E^\prime&\to&E&\to&{\cal O}_C(d+1)&\to &0\\
 &&\uparrow& &\uparrow&&\\
 &&E(-C)&=&E(-C)&\\
 &&\uparrow&&\uparrow&\\
 &&0&&0\\
\end{array}
\enq
We can easily obtain
\beqy
c_1(E^\prime)&=&c_1(E)-C=L-3C/2,\\
c_2(E^\prime)&=&c_2(E)+d=n+d.
\enqy
We remark that the correspondence between $(E,\varphi_{d+1})$
and $(E^\prime,\varphi_{-d})$ is bijective.
By using the above diagram, we have
\beq
\sum_{E\in M_{H_\infty}(L-3C/2,n)}\#{\tilde \Phi }_{-d}(E)
=\sum_{E\in M_{H_\infty}(L-C/2,n-d)}\#{\tilde \Phi }_{d+1}(E)
,(d\ge 0),
\enq
\beq
\sum_{E\in M_{H_\infty}(L-C/2,n)}\#{\tilde \Phi }_{-d}(E)
=\sum_{E\in M_{H_\infty}(L-3C/2,n-d-1)}\#{\tilde \Phi }_{d+1}(E),
(d\ge 0).
\enq
Using $M_{H_\infty}(L-C/2,n)_0({\bf F}_q)=\coprod_{l=0}^{n}{\tilde M}_{1,n}(l)$ 
and $M_{H_\infty}(L-3C/2,n)_0({\bf F}_q)=\coprod_{l=-1}^{n}{\tilde M}_{3,n}(l)$,
 we rewrite the above relations as follows:
\begin{lem}
\beq
\# {\tilde M}_{1,n}(d)
=\sum_{l=-1}^{d-3}q^{2d-2}(q^2-1)\# {\tilde M}_{3,n-d-1}(l)
+q^{2d}\# {\tilde M}_{3,n-d-1}(d-2),(d\ge 1),
\enq
\beq
\# {\tilde M}_{3,n}(d)
=\sum_{l=0}^{d-1}q^{2d}(q^2-1)\# {\tilde M}_{1,n-d}(l)
+q^{2d+2}\# {\tilde M}_{1,n-d}(d),(d\ge 0),
\enq
\beq
\# {\tilde M}_{1,n}(0)
=\# {\tilde M}_{3,n-1}(-1).
\enq
\end{lem}
By using these lemmas, we find
\beq
\# M_{H_\infty}(L-C/2,n)_0({\bf F}_q)
=\sum_{k=0}^nB_k^1(q)\# {\tilde M}_{1,n-k}(0),
\enq
\beq
\# M_{H_\infty}(L-3C/2,n)_0({\bf F}_q)
=\sum_{k=0}^{n+1}B_k^1(q)\# {\tilde M}_{1,n-k+1}(0).
\enq
We can arrange these formulas in the following form
\beqy
&&
\sum_n(\#M_{H_\infty}(L-C/2,n)_0({\bf F}_q)
+t\#M_{H_\infty}(L-3C/2,n)_0({\bf F}_q))t^n
\no
\\
&=&
2(\sum_nB_n^1(q)t^n)(\sum_n\#M_H(L-C/2,n)_0({\bf F}_q)t^n).
\enqy
$\sum_nB_n^1(q)t^n$ itself is given by
\begin{thm}
\beq
\sum_n B_n^1(q)t^n
=\prod_{d\ge 1}\frac{1-q^{4d-2}t^{2d}}{1-q^{4d-2}t^{2d-1}}.
\enq
\end{thm}

\subsection{{\cal O}(-2) Blow-up Formula}
To count the ${\bf{F}}_q$-rational points 
of the moduli space $M_{H_\infty}(L-iC/2,n)$,
we have to consider stable sheaves of non-vector bundle
(See Sec.5 in\cite{yoshioka}).
Let $E$ be a stable sheaf of rank $2$ on $\hat{X}$
with $c_1(E)=L-iC/2,c_2(E)=n$.
Then $E^{\vee\vee}$ is the double dual of $E$
and a stable vector bundle.
As mentioned above, there is an exact sequence:
\beq
0\to E \to  E^{\vee\vee} \to E^{\vee\vee}/E \to 0,
\enq
where $c_2(E^{\vee\vee})=c_2(E)-n^\prime,n^\prime=\dim_{{\bf{F}}_q}H^0(E^{\vee\vee}/E)$.
The number of stable sheaves with the above conditions
is $\#\mbox{Quot}_{E^{\vee\vee}/\hat{X}/{\bf{F}}_q}^{n^\prime}({\bf{F}}_q)$,
where the Quot-scheme Quot${}_{E^{\vee\vee}/\hat{X}/{\bf{F}}_q}^{n^\prime}$
is the scheme paremetrizing all quotients $\psi:E^{\vee\vee}\to A$
such that the Hilbert polynomial of $A$ is $n$.
Therefore $\#M_{H_\infty}(L-iC/2,n)({\bf{F}}_q)=\sum_{j=0}^{n}\#\mbox{Quot}_{E^{\vee\vee}/\hat{X}/{\bf{F}}_q}^{j}({\bf{F}}_q)\#M_{H_\infty}(L-iC/2,n-j)_0({\bf{F}}_q)$,
 and
\beq
\sum_n\#M_{H_\infty}(L-iC/2,n)({\bf{F}}_q)t^n=\left(\sum_{n}\#\mbox{Quot}_{{\cal O}_{\hat X}^{\oplus 2}/\hat{X}/{\bf{F}}_q}^{n}({\bf{F}}_q)\right)\left(\sum_n\#M_{H_\infty}(L-iC/2,n)_0({\bf{F}}_q)\right).
\enq
Here we use $E^{\vee\vee}\cong {\cal O}_{\hat X}^{\oplus 2}$.
This formula can be interpreted as the relation between
the uncompactified moduli spaces and the Gieseker moduli spaces
\cite{Li-Qin,Li-Qin2}.We quote the following useful formula. 
\begin{thm}(Yoshioka)\cite{yoshioka}
\beq
\sum_{n\ge 0}\#\mbox{\rm Quot}_{{\cal O}_{X}^{\oplus r}/{X}/{\bf{F}}_q}^{n}({\bf{F}}_q)=\prod_{a\ge 1}\prod_{b=1}^rZ_q(X,q^{ra-b}t^a).
\enq
\end{thm}
Since $\#{\hat X}({\bf F}_q)=\#{X}({\bf F}_q)+q$,we have
\beq
Z_q({\hat X},t)=Z_q(X,t)\times \frac{1}{1-qt}.
\enq
Substituting (4.38) and (4.21) into (4.36), we obtain 
\beqy
&&\sum_n(\# M_{H_\infty}(L,n)({\bf F}_q)
+t^{1/2}\# M_{H_\infty}(L-C,n)({\bf F}_q))t^n
\no
\\
=&&\frac{\sum_n (B_n^0(q)+t^{1/2}B_n^2(q))t^n}
{\prod_{a=1}(1-q^{2a-1}t^a)(1-q^{2a}t^a)}
\left(\sum_n(\# M_H(L,n)+t^{1/2}\# M_H(L-C,n))t^n\right).
\no\\&&
\enqy
Similarly for (4.33), we have
\beqy
&&
\sum_n(\#M_{H_\infty}(L-C/2,n)_0({\bf F}_q)
+t\#M_{H_\infty}(L-3C/2,n)_0({\bf F}_q))t^n
\no
\\
&=&
\frac{2\sum_nB_n^1(q)t^n}
{\prod_{a=1}(1-q^{2a-1}t^a)(1-q^{2a}t^a)}
\left(\sum_n\#M_H(L-C/2,n)t^n\right).
\enqy
The second factor of (4.39)((4.40)) is the generating function of 
Poincar\'e polynomials of the moduli spaces of stable sheaves on $X$.
We call this as the contribution from the untwisted sector.
On the other hand, the first factor of (4.39)((4.40))
is the contribution from the twisted sector.
This comes from the effect of the blow-up.
Thus we call this the ${\cal O}(-2)$ blow-up formula.

\subsection{$X={\bf C}^2/{\bf Z}_2$ Case}
Using (4.39), we derive the generating function
of Betti numbers of 
the moduli space of semistable sheaves on $\widehat{{\bf C}^2/{\bf Z}_2}$
(Note that if $c_1\equiv 0$ mod $2$,
we must consider semistable sheaves\cite{oko}). 
For this purpose, we have to derive Betti numbers of 
the moduli space of semistable sheaves on ${{\bf C}^2/{\bf Z}_2}$.
Considering that (4.36) is a universal relation,
we apply (4.36) to ${{\bf C}^2/{\bf Z}_2}$ case,
and separate this into the contribution from vector bundles
and that from sheaves of non-vector bundle.
To derive the contribution from vector bundles,
we consult the relation $({\bf C}^2/\Gamma)^{[n]}=(({\bf C}^2)^{[n]})^{\Gamma}$
\cite{nak},where $(*)^{\Gamma}$ stands for $\Gamma$-invariant part of $(*)$.
Although $(({\bf C}^2)^{[n]})^\Gamma=M(1,n)^\Gamma$ is the total contribution of rank $1$ case, we use the notion
that the moduli space of semistable vector bundles (sheaves) on ${\bf C}^2/\Gamma$
can be obtained by taking $\Gamma$-invariant part of 
the moduli space on ${\bf C}^2$.
First we derive the moduli space of semistable vector bundles on ${\bf C}^2$.
Using Theorem 11, we obtain
\beqy
\sum_{n\ge 0}\#\mbox{\rm Quot}_{{\cal O}_{{\bf C}^2}^{\oplus r}/{{\bf C}^2}/{\bf{F}}_q}^{n}({\bf{F}}_q)&=&\prod_{a\ge 1}\prod_{b=1}^rZ_q({\bf C}^2,q^{ra-b}t^a)
\no\\
&=&\frac{1}{\prod_{a\ge 1}\prod_{b=1}^r(1-q^{ra-b}t^a)}.
\enqy
Remembering the result of the framed moduli space of instanton
on $S^4$(for an example,Corrary 3.10 in\cite{n-y})
and comparing this formula with (4.36), we conclude 
\beq
\sum_n\#M_{H}(0,n)_0({\bf{F}}_q)=1,\sum_n\#M_{H}(-C,n)_0({\bf{F}}_q)=0.
\enq
Thus, as the moduli space of semistable vector bundles
on ${\bf C}^2/{\bf Z}_2$, we obtain
\beq
\sum_n\#M_{H}(0,n)_0^{{\bf Z}_2}({\bf{F}}_q)=1,
\sum_n\#M_{H}(-C,n)_0^{{\bf Z}_2}({\bf{F}}_q)=0,
\enq
where $M_{H}(0,0)_0^{{\bf Z}_2}\ni {\cal O}_{({\bf C}^2-\{\vec{0}\})/{\bf Z}_2}^{\oplus r}$ 
comes from $M_{H}(0,0)_0\ni {\cal O}_{{\bf C}^2}^{\oplus r}$.
Finally we obtain
\beqy
&&\sum_n(\# M_{H_\infty}(0,n)({\bf F}_q)
+t^{1/2}\# M_{H_\infty}(-C,n)({\bf F}_q))t^n
\no
\\
=&&\frac{\sum_n (B_n^0(q)+t^{1/2}B_n^2(q))t^n}
{\prod_{a=1}(1-q^{2a-1}t^a)(1-q^{2a}t^a)}
\cdot\frac{1}{{\prod_{a=1}(1-q^{2a-2}t^a)(1-q^{2a-1}t^a)}}.
\no\\&&
\enqy
By replacing $t\to q,q\to t^2$,
the right hand side is completely the same 
as the formula in the previous section.  
By replacing $c_2(E)=n\to c_2(E)-\frac{1}{4}c_1(E)^2$,
the left hand side also coincides with that in the previous section.
The equivalence between (3.64) and (4.44)
implies that the ${\cal O}(-2)$ blow-up formula is the true one.
Furthermore this implies that the calculation based 
on the fixed points of torus action is the same as
the calculation based on the fixed points of Frobenius morphism.

\section{Conclusion and Discussion}
\setcounter{equation}{0}
We derived the generating functions of Betti numbers of the framed moduli space of instantons on $\widehat{{\bf C}^2/{\bf Z}_2}$,
under the assumption that the corresponding torsion free sheaves
$E$ have vanishing properties($Hom(E,E(-l_\infty))=Ext^2(E,E(-l_\infty))=0$).
Combining Betti numbers of the moduli space of stable sheaves 
 on ${\bf C}^2/{\bf Z}_2$ with this,
we can obtain the ${\cal O}(-2)$ blow-up formula for rank two.
However this formula comes only from the calculation on ${\bf C}^2/{\bf Z}_2$
and that on  $\widehat{{\bf C}^2/{\bf Z}_2}$.
It is not clear that the ${\cal O}(-2)$ blow-up formula
is valid on any four surface with an $A_1$-singularity.

On the other hand, we derived  the ${\cal O}(-2)$ blow-up formula for rank two,
by determing the relation between 
the generating function of Betti numbers of the moduli space of 
stable sheaves on $\hat{X}$ and that on $X$.
We also checked this ${\cal O}(-2)$ blow-up formula
in $X={\bf C}^2/{\bf Z}_2$ case.
However the derivation was owing to the formal generalization 
of the ${\cal O}(-1)$ blow-up formula given by Yoshioka\cite{yoshioka}.
We did not consider the justification in ${\cal O}(-2)$ case in detail.

The above two methods to determine the ${\cal O}(-2)$ blow-up formula
have both advantages and disadvantages.
We assume that each method compensates each other.
The resulting ${\cal O}(-2)$ blow-up formula 
must be the true one
on any four surface with an $A_1$-singularity.
We also have to overcome the disadvantages in each side.
Can we derive vanishing properties($Hom(E,E(-l_\infty))=Ext^2(E,E(-l_\infty))=0$)
directly ?
Can we make mathematically more rigorous derivation
for the ${\cal O}(-2)$ blow-up formula ?
The most difficult point is the treatment 
of a stable sheaf on a singular surface.
One possibility to avoid this difficulty
is considering the contribution from the untwisted sector
instead of considering a stable sheaf on a singular surface.
Moreover we must define fractional line bundles ${\cal O}(k_\alpha C/2)$
rigorously. Indirectly fractional line bundles are justified
by the beautiful close formulas.

We want to estimate the tangent space 
$T_{(E,\Phi)}\hat{M}(2,2m+1,n)$ soon.
This completes the estimation of 
$T_{(E,\Phi)}\hat{M}(r,k,n)$ for general $r,k$,
which is described by the data of $T_{(E,\Phi)}\hat{M}(2,2m,n)$
and $T_{(E,\Phi)}\hat{M}(2,2m+1,n)$.
Then, we can derive Poincar\'e polynomials
of $\hat{M}(r,k,n)$ by using date of the tangent space.
According to the results of the $U(2)$ partition function
on $\widehat{{\bf C}^2/{\bf Z}_2}$,
we assume that the $U(r)$ partition function
on $\widehat{{\bf C}^2/{\bf Z}_2}$
is also described by a level 2 theta function $\theta_{A_{r-1}}^{(2)}$.
By considering the full $U(r)$ partition function,
we would find the duality between the $U(r)$ partition function on $\widehat{{\bf C}^2/{\bf Z}_2}$ and the $U(2)$ partition function on $\widehat{{\bf C}^2/{\bf Z}_r}$, which is concerned about
the level-rank duality of WZW models in conformal field theory\cite{nakanishi}.

We can consider instanton calculus on
$\widehat{{\bf C}^2/{\bf Z}_m}$.
Then, we assume that the $U(n)$ partition function
on $\widehat{{\bf C}^2/{\bf Z}_m}$
is also described by affine Lie algebras.  
We are interested in 
how the $U(n)$ partition function on $\widehat{{\bf C}^2/{\bf Z}_m}$
can be separated into the contribution from the twisted sector
and that from the untwisted sector.

We can also consider instanton calculus 
on $\widehat{{\bf C}^2/{\bf Z}_2}$ for $D,E$ gauge groups.
Then, we can use the duality between 
the  $D,E$ partition function on $\widehat{{\bf C}^2/{\bf Z}_2}$
and the  $U(2)$ partition function on $\widehat{{\bf C}^2/\Gamma_{D,E}}$.
Furthermore we can derive the corresponding blow-up formula.
These calculations must serve to a further insight 
into our works for $D,E$ gauge groups\cite{jin3,sasaki1,sasaki2,sasaki3}.

What is the meaning of the middle factor in (3.63) ?
Since this factor does not come from affine Lie algebras,
it must be complicated to derive the generating functions
for $D,E$ gauge groups.

We want to verify the equivalence between the generating function
on orbifold $T^4/{\bf Z}_2$ and that on $K3$,
by using the ${\cal O}(-2)$ blow-up formula derived in this article.
Then, we are interested in how 16th power of the ${\cal O}(-2)$ blow-up formula
are combined with the untwisted sector of the generating function
on orbifold $T^4/{\bf Z}_2$.
We hope to verify the equivalence by not assuming the duality conjecture.

{\bf Acknowledgment}\\
We would like to thank H.Awata,S.Fujii, T.Hausel, 
M.Jinzenji, H.Kanno, S.Kondo, S.Minabe,  
and K.Yoshioka for helpful suggestions
and useful discussions.

\appendix
\renewcommand{\theequation}{\alph{section}.\arabic{equation}}
\section{Theta Functions}
\subsection{Level $l$ Theta Functions}
Following \cite{kac},
a level $l~A_r$ theta function for weight $\beta$ is given by 
\beq
\theta_{A_{r}}^{\beta(l)}(\tau;y):=\sum_{m\in{\bf Z}^{r}}q^{\frac{l}{2}{}^t(m+\frac{1}{l}A_{r}^{-1}\beta)A_{r}(m+\frac{1}{l}A_{r}^{-1}\beta)}
e^{2\pi i l{}^t(m+\frac{1}{l}A_{r}^{-1}\beta)A_{r}y},{}^t\beta=(b_1,b_2,\ldots,b_{N-1}),b_i\in {\bf Z},y\in {\bf Z}^r.
\enq
Here $A_r$ stands for the Cartan matrix of $A_r$ again.
$q=\exp(2\pi i \tau)$.

Using this notation, we give Jacobi's theta functions:
\beq
\theta_3(\tau;y)=\theta_{A_1}^{0(2)}(\tau;y)+\theta_{A_1}^{2(2)}(\tau;y),
\enq
\beq
\theta_4(\tau;y)=\theta_{A_1}^{0(2)}(\tau;y)-\theta_{A_1}^{2(2)}(\tau;y),
\enq
\beq
\theta_2(\tau;y)=\theta_{A_1}^{1(2)}(\tau;y).
\enq
Similarly we define $\theta_{A_r}^{(2)}(\tau;y)$ as
\beq
\theta_{A_r}^{(2)}(\tau;y):=\sum_{{}^t(A_r\beta)=(\epsilon_1,\epsilon_2,\ldots,\epsilon_r)\atop \epsilon_k=0~ or~ 1}
\theta_{A_r}^{\beta(2)}(\tau;y).
\enq
Using $q=\exp(2\pi i)$ and $z=\exp(2\pi i y)$, we sometimes
denote these theta functions by
$\theta_{A_r}^{(2)}(q;y),\theta_{A_r}^{(2)}(\tau;z)$ 
and $\theta_{A_r}^{(2)}(q;z)$.
Note that $\theta_{A_r}^{(2)}(\tau;y=0)$ is invariant under $\tau\to -\frac{1}{\tau}$ up to a factor.

\subsection{Level $2$ Theta Functions via Instanton Calculus on $\widehat{{\bf C}^2/{\bf Z}_2}$}
First we derive the constraint (3.52) 
from $E=\sum_\alpha I_\alpha\otimes {\cal O}(k_\alpha C/2)$.
We have $c_1(E)=\sum_\alpha k_\alpha C/2, c_2(E)=\sum_\alpha|Y_\alpha|
+C^2\sum_{\alpha<\beta}k_\alpha k_\beta/4$ at the fixed point.
Then we obtain
\beqy
c_2(E)-\frac{r-1}{2r}c_1(E)^2
&=&\sum_\alpha|Y_\alpha|+C^2\sum_{\alpha<\beta}\frac{k_\alpha k_\beta}{4}
-\frac{r-1}{2r}C^2\frac{(\sum_\alpha k_\alpha)^2}{4}
\no
\\
&=&
\sum_\alpha|Y_\alpha|+\frac{1}{r}\sum_{\alpha<\beta}|\frac{k_\alpha}{2}-\frac{k_\beta}{2}|^2.
\enqy
Under the condition $\sum_\alpha k_\alpha=k$,
we show that $\sum_{k_\alpha\in{\bf Z}}q^{\frac{1}{r}\sum_{\alpha<\beta}|\frac{k_\alpha}{2}-\frac{k_\beta}{2}|^2}$ becomes level $2$ theta functions.
By reparametrizing $\sum_{\alpha=i}^rk_\alpha=K_i$, we obtain
\beqy
\frac{1}{r}\sum_{\alpha<\beta}|\frac{k_\alpha}{2}-\frac{k_\beta}{2}|^2
&=&-\frac{1}{2}\sum_{\alpha<\beta}k_\alpha k_\beta+\frac{r-1}{r}k^2
\no\\
&=&\frac{1}{2}(K_2-k)K_2+\frac{1}{2}\sum_{i=3}^r(K_i-K_{i-1})K_i+\frac{r-1}{r}k^2
\no\\
&=&
{}^t(\frac{K}{2}-\frac{k}{2}A_{r-1}^{-1}v)A_{r-1}
(\frac{K}{2}-\frac{k}{2}A_{r-1}^{-1}v),
\enqy 
where $K\in {\bf Z}^{r-1},{}^tv=(1,0,..,0)$.
This is nothing but the exponent of (a.1) for $l=2$ case.
Conversely if $k_\alpha/2$ is replaced by $k_\alpha$ in (a.7),
we would not obtain the level $2$ theta functions.

\setcounter{equation}{0}
\section{The Proof of the Equality between (3.62) and (3.63)}
\setcounter{equation}{0}
 Let
 \beq
(a)_k=(a;q)_k=(1-a)(1-aq)\cdots(1-aq^{k-1}),
(a)_\infty=\prod_{d=0}^{\infty}(1-aq^d).
 \enq
We start with the formula of Entry 7(p.16) in \cite{ram}.
We substitute $a=b$,then we obtain
\beq
\sum_{k\ge 0}^{\infty}\frac{(d/c)_k(d/q)_k(1-dq^{2k-1})c^kq^{k(k-1)}}
{(c)_k(q)_k}=\frac{(d/q)_\infty}{(c)_\infty}.
\enq
We put $c=-q,d=q^2a$,
\beq
\sum_{k\ge 0}^{\infty}\frac{(-qa)_k(qa)_k(1-aq^{2k+1})(-1)^kq^{k^2}}
{(-q)_k(q)_k}=\frac{(qa)_\infty}{(-q)_\infty}.
\enq
Using $(e)_k(-e)_k=(e^2;q^2)_k$,
\beqy
&&\sum_{k\ge 0}^\infty\left(
\frac{(q^2a^2;q^2)_k}{(q^2;q^2)_k}(-q)^{k^2}
+\frac{(q^2a^2;q^2)_k}{(q^2;q^2)_k}(-q)^{(k+1)^2}a
\right)
\no\\
&=&
\frac{(qa)_\infty}{(-q)_\infty}.
\enqy
We set $q=-t^2q^{\frac{1}{2}},a=t^{-2}$,
\beqy
&&\sum_{k\ge 0}^\infty\left(
\prod_{d=1}^k\frac{1-t^{4d-4}q^d}{1-q^{4d}q^d}t^{2k^2}q^{\frac{k^2}{2}}
+\prod_{d=1}^{k-1}\frac{1-t^{4d-4}q^d}{1-q^{4d}q^d}
t^{2k^2-2}q^{\frac{k^2}{2}}
\right)
\no\\
&=&
\prod_{d=1}^\infty\frac{1-(-t^2q^{\frac{1}{2}})^dt^{-2}}
{1+(-t^2q^{\frac{1}{2}})^d}.
\enqy
We verify the equality between (c.5) and (c.6).
We put $c=q^{\frac{3}{2}},d=aq^2$ for Entry 7,and obtain
\beq
\sum_{k\ge 0}^{\infty}\frac{(aq^{\frac{1}{2}})_k(aq)_k(1-aq^{2k+1})
q^{k^2+\frac{k}{2}}}
{(q^{\frac{3}{2}})_k(q)_k}=\frac{(aq)_\infty}{(q^{\frac{3}{2}})_\infty}.
\enq
Using $(e)_k(eq^{\frac{1}{2}})_k=(e;q^{\frac{1}{2}})_{2k}$,
\beq
\sum_{k\ge 0}^\infty
\frac{(aq^{\frac{1}{2}};q^{\frac{1}{2}})_{2k}(1-aq^{2k+1})}
{(q^{\frac{1}{2}};q^{\frac{1}{2}})_{2k}(1-q^{k+\frac{1}{2}})}q^{k^2+\frac{k}{2}}=
\frac{(aq)_\infty}{(q^{\frac{1}{2}})_\infty}.
\enq
\beq
\sum_{k\ge 0}^\infty
\frac{(aq^{\frac{1}{2}};q^{\frac{1}{2}})_{2k}}
{(q^{\frac{1}{2}};q^{\frac{1}{2}})_{2k}}q^{k^2+\frac{k}{2}}
\left(1+\frac{1-aq^{k+\frac{1}{2}}}{1-q^{k+\frac{1}{2}}}q^{k+\frac{1}{2}}\right)
=
\frac{(aq)_\infty}{(q^{\frac{1}{2}})_\infty}.
\enq
\beq
\sum_{k\ge 0}^\infty\left(
\frac{(aq^{\frac{1}{2}};q^{\frac{1}{2}})_{2k}}
{(q^{\frac{1}{2}};q^{\frac{1}{2}})_{2k}}q^{\frac{(2k)^2+2k}{4}}
+\frac{(aq^{\frac{1}{2}};q^{\frac{1}{2}})_{2k+1}}
{(q^{\frac{1}{2}};q^{\frac{1}{2}})_{2k+1}}
q^{\frac{(2k+1)^2+2k+1}{4}}\right)
=
\frac{(aq)_\infty}{(q^{\frac{1}{2}})_\infty}.
\enq
\beq
\sum_{k\ge 0}^\infty
\frac{(aq^{\frac{1}{2}};q^{\frac{1}{2}})_{k}}
{(q^{\frac{1}{2}};q^{\frac{1}{2}})_{k}}q^{\frac{k^2+k}{4}}
=
\frac{(aq)_\infty}{(q^{\frac{1}{2}})_\infty}.
\enq
We set $q=t^8q^{2},a=t^{-4}$,
\beq
\sum_{k\ge 0}^\infty\prod_{d=1}^k\frac{1-t^{4d-4}q^d}{1-t^{4d}q^d}
t^{2k^2+2k}q^{\frac{k^2+k}{2}}
=\prod_{d=1}^\infty\frac{1-t^{8d-4}q^{2d}}{1-t^{8d-4}q^{2d-1}}.
\enq

\section{Betti Numbers of ${\hat M}(r,2m+1,n)$}
\setcounter{equation}{0}
As in the proof of the theorem 6,for odd $n$ 
$H^1(F_2,{\cal O}(-nC/2-l_\infty))$ is not equal to
\beq
\bigoplus_{d=0}^{(n-1)/2-1}H^0({\bf P}^1,{\cal O}_{{\bf P}^1}(2d+1)).
\enq
However by using this to count indices, we obtain the following formulas.

The poincar\'e polynomial of ${\hat M}(2,2m+1,n)$ is given by
\begin{equation}
P_t({\hat M}(2,2m+1,n))=\sum \prod_{\alpha=1}^2 t^{2(|Y_\alpha|-l(Y_\alpha))}
P_t(S^{Y_\alpha}{\bf P}^1)
\cdot t^{2(l^\prime+|Y_1|+|Y_2|-n^\prime)},
\end{equation}
where $k\equiv 1$ case,
\begin{equation}
l^\prime=\left\{
\begin{array}{c}
(K+m)^2+(K+m) \mbox{ if }k_1=K+2m+1> k_2=-K,\\
(-K-m)^2+(K+m) \mbox{ if } k_1=K+2m+1 < k_2=-K,
\end{array}
\right.
\end{equation}
\begin{equation}
n^\prime=\left\{
\begin{array}{c}
(\# \mbox{ of columns of $Y_1$ which are longer than $K+m)$ if $K+m\ge 0$},\\
(\#  \mbox{ of columns of $Y_2$ which are longer than $-K-m-1)$ if $K+m < 0$}.
\end{array}
\right.
\end{equation}

The generating function of Poincar\'e polynomial in the case of $r=2,c_1=1$ is
\begin{eqnarray}
&&\sum_{n\in {{\bf Z}\ge 0}}P_t({\hat M}(2,1,n+\frac{1}{8}))q^{n+\frac{1}{8}}
\nonumber \\
=&&
\prod_{d=1}^\infty\frac{1}{(1-t^{4d}q^d)(1-t^{4d-2}q^d)^2(1-t^{4d-4}q^d)}
\nonumber \\
&& \times \left[\sum_{k\ge 0}\prod_{d=1}^k\frac{1-t^{4d-4}q^d}{1-t^{4d}q^d}
t^{2k^2+2k}q^{\frac{(k+\frac{1}{2})^2}{2}}
+\sum_{k > 0}\prod_{d=1}^{k-1}\frac{1-t^{4d-4}q^d}{1-t^{4d}q^d}
t^{2k^2-2k}q^{\frac{(k-\frac{1}{2})^2}{2}}
\right]
\no\\
\\
=&&
\prod_{d=1}^\infty\frac{1}{(1-t^{4d}q^d)(1-t^{4d-2}q^d)^2(1-t^{4d-4}q^d)}
\prod_{d=1} \frac{1-t^{8d-4}q^{2d}}{1-t^{8d}q^{2d}}
\sum_{k\in{\bf Z}}t^{2k^2+2k}q^{\frac{(k+\frac{1}{2})^2}{2}}
\no\\&&
\\
=&&
2q^{\frac{1}{8}}\prod_{d=1}^\infty\frac{1}{(1-t^{4d}q^d)(1-t^{4d-2}q^d)^2(1-t^{4d-4}q^d)}
\prod_{d=1}\frac{1-t^{8d-4}q^{2d}}{1-t^{8d-4}q^{2d-1}}.
\no\\
\end{eqnarray}
This result is consistent with (4.40).

\end{document}